 \documentstyle[equations,epsfig,11pt]{article}

 \renewcommand{\theequation}{\arabic{equation}}
\begin{document}
 \bibliographystyle{plain}
 \def\m@th{\mathsurround=0pt}
 \mathchardef\bracell="0365
 \def\upbrall{$\m@th\bracell$}
 \def\undertilde#1{\mathop{\vtop{\ialign{##\crcr
     $\hfil\displaystyle{#1}\hfil$\crcr
      \noalign
      {\kern1.5pt\nointerlineskip}
      \upbrall\crcr\noalign{\kern1pt
    }}}}\limits}
 \def\theequation{\arabic{section}.\arabic{equation}}
 \newcommand{\ar}{\alpha}
 \newcommand{\aar}{\bar{a}}
 \newcommand{\bb}{\beta}
 \newcommand{\gm}{\gamma}
 \newcommand{\Gm}{\Gamma}
 \newcommand{\en}{\epsilon}
 \newcommand{\dd}{\delta}
 \newcommand{\sg}{\sigma}
 \newcommand{\kp}{\kappa}
 \newcommand{\ld}{\lambda}
 \newcommand{\oa}{\omega}
 \newcommand{\Ups}{\Upsilon}
 \newcommand{\be}{\begin{equation}}
 \newcommand{\ee}{\end{equation}}
 \newcommand{\bea}{\begin{eqnarray}}
 \newcommand{\eea}{\end{eqnarray}}
 \newcommand{\bse}{\begin{subequations}}
 \newcommand{\ese}{\end{subequations}}
 \newcommand{\nn}{\nonumber}
 \newcommand{\bR}{\bar{R}}
 \newcommand{\bP}{\bar{\Phi}}
 \newcommand{\bS}{\bar{S}}
 \newcommand{\bU}{\bf U}
 \newcommand{\bW}{\bar{W}}
 \newcommand{\vf}{\varphi}
 \newcommand{\sn}{{\rm sn}}
 \newcommand{\cn}{{\rm cn}}
 \newcommand{\dn}{{\rm dn}}
 \newcommand{\pl}{\partial}
 \newcommand{\ddp}{\frac{\partial}{\partial p}}
 \newcommand{\ddq}{\frac{\partial}{\partial q}}
 \newcommand{\wh}{\widehat}
 \newcommand{\ol}{\overline}
 \newcommand{\wt}{\widetilde}
 \newcommand{\ut}{\undertilde}
 \newcommand{\Ld}{{\bf \Lambda}}
 \newcommand{\tLd}{\,^{t\!}{\bf \Lambda}}
 \newcommand{\I}{{\bf I}}
 \newcommand{\tII}{\,^{t\!}{\bf I}}
 \newcommand{\tuk}{\,^{t\!}{\bf u}_k}
 \newcommand{\tul}{\,^{t\!}{\bf u}_\ell}
 \newcommand{\tcl}{\,^{t\!}{\bf c}_{\ell}}
 \newcommand{\ssk}{\sigma_{k^\prime}}
 \newcommand{\ssl}{\sigma_{\ell^\prime}}
 \newcommand{\ddint}{\int_\Gamma d\ld(\ell) }
 \def\hypotilde#1#2{\vrule depth #1 pt width 0pt{\smash{{\mathop{#2}
 \limits_{\displaystyle\widetilde{}}}}}}
 \def\hypohat#1#2{\vrule depth #1 pt width 0pt{\smash{{\mathop{#2}
 \limits_{\displaystyle\widehat{}}}}}}
 \def\hypo#1#2{\vrule depth #1 pt width 0pt{\smash{{\mathop{#2}
 \limits_{\displaystyle{}}}}}}

 \begin{center}
 {\large{\bf On Discrete Painlev\'e Equations\\
 Associated with the Lattice KdV Systems\\
 and the Painlev\'e VI Equation}}
 \vspace{1.2cm}
 
 F.W. Nijhoff\vspace{.15cm} \\
 {\it Department of Applied Mathematical Studies\\
 The University of Leeds, Leeds LS2 9JT, UK}\\
 \vspace{.3cm}
 A. Ramani\vspace{.15cm}\\
 {\it CPT, Ecole Polytechnique\\
 CNRS, UPR 14\\
 91128 Palaiseau, France}\\
 \vspace{.2cm}
 B.Grammaticos\vspace{.15cm}\\
 {\it GMPIB, Universit\'e Paris 7\\
 2 Place Jussieu, Tour 24--14, 5eme \'etage\\
 75251 Paris, France}\\
 \vspace{.2cm}
 Y. Ohta\vspace{.15cm}\\
 {\it Department of Applied Mathematics\\
 Faculty of Engineering, Hiroshima University\\
 1-4-1 Kagamiyama, Higashi-Hiroshima, 739 Japan} \\
 \vspace{.4cm}
 \end{center}
 \vspace{1.4cm}
 \pagebreak
 
 \centerline{\bf Abstract}
 \vspace{.2cm}
 A new integrable nonautonomous nonlinear ordinary difference
 equation is presented which can be considered to be a discrete
 analogue of the Painlev\'e V equation. Its derivation is based on the
 similarity reduction on the two-dimensional lattice of integrable
 partial difference equations of KdV type. The new equation
 which is referred to as GDP (generalised discrete Painlev\'e
 equation) contains various ``discrete Painlev\'e equations''
 as subcases for special values/limits of the parameters,
 some of which were already given before in the literature. The
 general solution of the GDP can be expressed in
 terms of Painlev\'e VI (PVI) transcendents. In fact, continuous PVI
 emerges as the equation obeyed by the solutions of the discrete
 equation in terms of the lattice {\it parameters} rather than the
 lattice {\it variables} that label the lattice sites. We show that
 the bilinear form of PVI is embedded naturally
 in the lattice systems leading to the GDP.
 Further results include the establishment of B\"acklund and
 Schlesinger transformations for the GDP, the
 corresponding isomonodromic deformation problem, and
 the self-duality of its bilinear scheme.
 \vfill
 
 
 \setcounter{page}{0}
 \pagebreak

 \section{Introduction}
 \setcounter{equation}{0}
 
 Discrete Painlev\'e equations \cite{Carg} have now been around for
 already a
 couple of years and they form a new exciting class of equations,
 possibly leading to the definition of new transcendental functions.
 Their study may yield new insights into the analytic theory of ordinary
 difference equations, the development of which, cf. \cite{Norlund},
 has been greatly lagging behind compared to its
 continuous counterpart --the theory of ordinary differential
 equations. In view of this, the recent interest in the investigation of
 {\it integrable} discrete systems may open a
 new window on what might ultimately become a general theory
 of discrete and difference equations.  What the study of integrable
 discrete systems has provided so far is a rich class of interesting
 examples of nonlinear ordinary and partial difference equations that
 are amenable to exact and rigorous approaches for their solution. With
 these examples we expect that insight can be gained into the general
 theory, and that they will help us to develop methods for the
 study not only of
 integrable equations but of difference equations in general.
 Integrable partial difference equations have been investigated
 over the last two decades, cf. \cite{AL}-\cite{QNCL}, cf. \cite{KDV}
 for a review, providing a large reservoir of examples for the study
 of further properties. To this many novel results and important
 insights have been added especially during the last few years
 comprising of: the study of initial value problems on the space-time
 lattice, \cite{PNC}, the study of similarity reductions on the
 lattice, \cite{NP}, the  establishment of the remarkable singularity
 confinement property of integrable discrete systems, \cite{GRP},
 i.e. an effective integrability detector for discrete systems
 (deemed to be an initial step towards finding a purely discrete
 analogue of the Painlev\'e property), the elaboration of proper
 discrete analogues of the Painlev\'e transcendental equations
 PI-PVI, \cite{NP,RGH,FIK,FGR}, the construction of special solutions
 and properties of the discrete Painlev\'e equations, (\cite{Carg}
 for a review).
 
 Let us provide a few examples. The discrete Painlev\'e I
 (dPI) equation
 \be
 x_{n+1}+x_n+x_{n-1}=\frac{\zeta_n}{x_n}+a \   \ ,\    \
 \zeta_n\equiv \ar n+\bb\   ,
 \label{eq:dPI}
 \ee
 arose from the theory of orthogonal polynomials (cf. \cite{Magnus}
 for a review),  and in physics in the matrix model approach to
 two-dimensional quantum gravity, \cite{BIZ}, cf. also \cite{FIK}.
 The discrete Painlev\'e II (dPII) equation
 \be
 x_{n+1}+x_{n-1}= \frac{\zeta_nx_n+a}{1-x_n^2}
 \label{eq:dPII}
 \ee
 was found in connection with unitary matrix models of quantum gravity
 in \cite{Per} and from the lattice similarity
 approach introduced in \cite{NP}. Other discrete Painlev\'e
 equations  have been found from
 the B\"acklund and Schlesinger transformations of the continuous
 Painlev\'e equations, cf. \cite{FGR}, for example the
 alternative discrete Painlev\'e II (alt-dPII) equation
 \be\label{eq:alt-dPII}
 \frac{\zeta_{n+1}}{x_{n+1}x_n+1} +
 \frac{\zeta_n}{x_nx_{n-1}+1} = \zeta_n - x_n+\frac{1}{x_n} + a\  ,
 \ee
 that was studied more extensively in \cite{altdp2}. Many of
 these discrete Painlev\'e equations have been shown to possess
 an associated isomonodromic deformation problem, cf.
 \cite{PNGR}-\cite{JimboS}, (with the notable exception of the
 discrete dPIV and dPV equations found in \cite{RGH},
 for which so far no monodromy problem has been established).
 Thus, we have now at our disposal a large body of examples
 of discrete equations that can be considered to be the proper analogues
 of the continuous equations. There does not yet, of course, exist a
 full-fledged theory on the discrete level, but already many of the
 outstanding features of discrete Painlev\'e equations
 have been established, such as the existence of
 special solutions for special values of the parameters, the associated
 bilinear forms, the B\"acklund/Schlesinger transformations, etc.
 (For a recent review, cf. \cite{Carg}).
 
 Central in this paper is a system of {\it partial} difference equations
 which can be thought of as a (scaling) symmetry reduction of the
 lattice MKdV equation, cf. \cite{NP,DIGP}. It consists of the following
 coupled set of equations:
 \bse\label{eq:MKdVsyst}
 \be \label{eq:dMKdV}
 p\left( \frac{v_{n,m+1}}{v_{n+1,m+1}}-\frac{v_{n+1,m}}{v_{n,m}}\right)
 = q\left( \frac{v_{n+1,m}}{v_{n+1,m+1}}-\frac{v_{n,m+1}}{v_{n,m}}
 \right) \   ,
 \ee
 which is the lattice MKdV equation, together with
 \be\label{eq:MKdVconstr}
 n\frac{v_{n+1,m}-v_{n-1,m}}{v_{n+1,m}+v_{n-1,m}} +
 m\frac{v_{n,m+1}-v_{n,m-1}}{v_{n,m+1}+v_{n,m-1}}  = \mu
 -\ld (-1)^{n+m} \   ,
 \ee \ese
 which is a nonlinear nonautonomous constraint expressing the scaling
 symmetry. The similarity constraint (\ref{eq:MKdVconstr})
 was found already some years ago in \cite{NP} for the special values
 of the parameters $\ld=0$, $\mu=0$. Related systems were developed
 recently, \cite{DIGP}, and a
 geometrical interpretation was given in \cite{Bobenko2}.
 
 It is shown that eqs. (\ref{eq:dMKdV}) and
 (\ref{eq:MKdVconstr}) are compatible on the lattice and that their
 reduction gives rise to a combined set of ordinary nonlinear
 nonautonomous difference equations of the form
 \bea\label{eq:MPII}
 && \frac{2\zeta_{n+1}}{1-X_{n+1}X_{n}} +
 \frac{2\zeta_{n}}{1-X_{n}X_{n-1}} = 
 \mu+\nu+ \zeta_{n+1}+\zeta_n + \nn \\
 &+& 
 \frac{ (\mu-\nu)(r^2-1)X_n + r(1-X_n^2)
 \left[ \frac{1}{2}(\zeta_n+\zeta_{n+1}) + \frac{1}{2}
 (-1)^n(\zeta_n-\zeta_{n+1}-2m) \right] }
 { (r+X_{n})(1+rX_{n})}\   ,
 \eea
 with parameters $r=q/p$, $m$, $\mu$ and the alternating parameter
 $\nu=\ld(-1)^{n+m}$, including the discrete-time step.
 We will refer to this discrete equation as GDP (generalised discrete
 Painlev\'e equation), since as we will show in what follows this
 second-order nonlinear nonautonomous difference equation incorporates many
 of the so
 far known discrete Painlev\'e equations of {\it difference}
 type\footnote{
 There exist also discrete Painlev\'e equations of $q$-difference
 type, such as the discrete PIII equation equation found first in
 \cite{RGH} from the singularity confinement analysis and for which
 a $q$-difference monodromy problem was derived in \cite{PNGR}.
 Further $q$-difference dP's include discrete PV, \cite{RGH} and
 discrete PVI found in \cite{JimboS}.}. In fact,
 special limits of (\ref{eq:MPII}) reduce to both discrete
 versions of PII, namely (\ref{eq:dPII}) and (\ref{eq:alt-dPII}).
 However, by itself, eq.
 (\ref{eq:MPII}) can be considered to be a discrete analogue of the
 PV equation by virtue of the fact that, as we shall show below,
 eq. (\ref{eq:MPII}) reduces to PV by a continuum limit. Thus,
 eq. (\ref{eq:MPII}) seems to be a very rich equation, and this
 point is made even stronger by noting that solutions of
 (\ref{eq:MPII}) can actually be expressed in terms of PVI
 transcendents, i.e. solutions of the PVI equation
 \bea\label{eq:PVI}
 \frac{d^2w}{dt^2}=&& \frac{1}{2}\left( \frac{1}{w}+\frac{1}{w-1}+
 \frac{1}{w-t}\right)\left(\frac{dw}{dt}\right)^2 -
 \left(\frac{1}{t}+\frac{1}{t-1}+\frac{1}{w-t}\right)
 \frac{dw}{dt}  \nn \\
 && + \frac{w(w-1)(w-t)}{t^2(t-1)^2}\left( \ar-\bb\frac{t}{w^2}
 +\gm\frac{t-1}{(w-1)^2}-(\dd-\frac{1}{2})\frac{t(t-1)}{(w-t)^2}\right)\   .
 \eea
 An exhaustive list of Schlesinger transformations for PVI has been
 given by Mugan and Sakka in \cite{Mugan}, and in
 principle the GDP equation (\ref{eq:MPII}) can be considered
 to be the permutability condition for the corresponding B\"acklund
 transformations in the spirit of \cite{FGR}. Also in the early
 paper \cite{JimboMiwa} difference relations associated with
 the Painlev\'e equations were considered, but no explicit closed-form
 discrete equations associated with PVI was given. The GDP
 (\ref{eq:MPII}) is, in fact, the appropriate difference equation
 that is naturally associated with PVI and the aim of the
 present paper is to study this equation, as an integrable
 ordinary difference equation in its own right.
 
 Let us point out that referring to equations like (\ref{eq:MPII})
 as a discrete PV (dPV) equation risks to become slightly confusing.
 Since, on the one hand, there are various different discrete
 Painlev\'e equations that tend to one and the same continuous
 Painlev\'e equation in limits, on the other hand the same discrete
 equation may have various limits to different continuous and/or
 discrete equations, this type of nomenclature (referring to discrete
 Painlev\'e equations in terms of the continuous PI-PVI equations
 to which they constitute the discrete analogues) is slightly
 deceptive. In particular, there exist various alternative ``dPV''
 equations as well and so far the interconnection between these
 various discrete equations is far from clear. Issues of
 classification and universality will not be addressed in this
 paper, but we hope that revealing the structure behind equations
 such as (\ref{eq:MPII}) might be of help in establishing the
 connections between various integrable ordinary difference
 equations.
 
 In this paper, we will study the lattice system (\ref{eq:MKdVsyst})
 as well as the GDP equation (\ref{eq:MPII}) from various angles.
 The outline of the paper is as follows. In section 2 we will present
 a general formalism in terms of an infinite matrix structure from which
 one can derive the lattice equations as well as their similarity
 constraints.  The explicit forms of the lattice systems are presented
 in section 3, together with their general  solution in terms of PVI
 transcendents. In section 4 we discuss the integrability
 of the lattice systems and their consistency from the point of view
 of initial value problems. Furthermore, from the structure developed
 in section 2, we derive their isomonodromic deformation problems.
 Finally, in section 5 we discuss the lattice equations from a general
 point of view studying various properties, such as their B\"acklund
 and Schlesinger transformations
 as well as the general bilinear scheme underlying these equations.
 In fact, writing the lattice equations in bilinear form we are led to a
 system of four-dimensional discrete equations in which parameters and
 variables live on equal footing. This self-duality of the bilinear forms,
 the so-called {\it Grand Scheme} of \cite{ROSG} where it was observed
 in another context, seems quintessential to the integrability
 of the lattice system.

 \section{Derivation: Infinite-Matrix Scheme}
 \setcounter{equation}{0}
 
 We start by developing a general scheme from which we obtain the
 discrete equations and their similarity reduction.
 The scheme is based on a structure introducing a dynamics in
 terms of infinite matrices, and it was used in the past to construct
 integrable partial difference equations together with their Lax pairs,
 Miura- and B\"acklund transformations and to establish the
 underlying algebraic structure, cf. \cite{NQC,QNCL}, cf. also
 the more recent account \cite{KDV} for a review. The scheme is
 formal, and we would like to stress that the objective of this
 derivation is exclusively to establish the basic equations and to
 unravel the underlying algebraic structure. We do not make any
 claim on the analytic aspects of the scheme: in order for it to be
 used to construct solutions of the equations one needs to establish
 the existence of the objects under question beyond the formal level.
 However, the scheme has proven to be highly effective in arriving at
 all the necessary relations between the equations which then in
 the next stage can be used for constructing analytic solutions.

 \subsection{Definitions}
 
 For brevity, let us formulate the structure in a rather
 abstract way. We basically need three ingredients to formulate the
 entire scheme:
 \begin{itemize}
 
 \item An infinite ( ${\bf Z}\times {\bf Z}$) matrix ${\bf C}$  which we
 can take of the form
 \be\label{eq:C}   {\bf C}=\ddint\,\rho_\ell {\bf c}_\ell\tcl\   , \ee
 in which ${\bf c}_\ell$ and $\tcl$ are infinite vectors with components
 $({\bf c}_\ell)_j=(\tcl)_j=\ell^j$, and $\rho_\ell$ depends on additional
 variables that are to be determined later. The integrations over
 contour $\Gamma$ and measure $d\ld$ need not be specified at this
 point but we we will loosely assume that they can be chosen such that the
 objects to be introduced below are well-defined. From the definition it is
 clear that ${\bf C}$ is symmetric: ~${\bf C}=\,^{t\!}{\bf C}$~,
 (the left superscript $t$ denoting the adjoint of the ${\bf Z}\times
 {\bf Z}$ matrix).
 \item
 Matrices $\Ld$ and $\tLd$ that define the operations of index-raising
 when multiplied at the left respectively the right, as well as
 matrices $\I$ and $\tII$ that count the index label, namely by
 $(\I\cdot{\bf c}_\ell)_j=j({\bf c}_\ell)_j$, $(\tcl\cdot\tII)_j=j(\tcl)_j$.
 The interrelation between the variables $\Ld$,$\tLd$,${\bf I}$ and
 $\tII$ is given by
 \bse \label{eq:interrels}\bea
 \Ld^j\cdot {\bf I}&=&({\bf I}+j)\cdot \Ld^j\   ,  \\
 \tII\cdot \tLd^j &=&\tLd\cdot (\tII +j)\   .
 \eea\ese
 \item
 An infinite matrix ${\bf \Omega}$ obeying the equations:
 \bse\label{eq:Omega}\bea
 {\bf \Omega}\Ld^j - (-\tLd)^j{\bf \Omega}={\bf O}_j\     , \\
 \tII\cdot {\bf \Omega} + {\bf \Omega}\cdot {\bf I}+{\bf \Omega}= 0\    .
 \eea\ese
 in which
 \be\label{eq:O}
 {\bf O}_k=\sum_{j=0}^{k-1} (-\tLd)^j\cdot{\bf O}\cdot\Ld^{k-1-j}\  ,
 \ee
 and where ${\bf O}$ is a projection matrix on the central element, i.e.
 $({\bf O}\cdot {\bf C}_k)_{i,j}=\dd_{i,0}{\bf C}_{0,j}$, etc. ,
 obeying also
 \be \label{eq:IO}
 \tII\cdot{\bf O}={\bf O}\cdot {\bf I}=0\   .
 \ee
 \end{itemize}
 
 There are linear  equations for the object ${\bf C}$, the form of
 which depends  on the choice of the factor $\rho_\ell$. We can choose
 $\rho_\ell$ to depend on continous time-variables
 ~$t_1,t_2,t_3,\dots $~, in terms of which we have linear evolutionary
 flows. However, since we deal here with the  discrete case we choose
 $\rho_\ell$ of the form\footnote{Throughout this paper we will
 use independent discrete variables $n_\nu$, or in particular
 $n$ and $m$. Strictly speaking, these variables need not
 be integers themselves, we only require that they {\it shift by
 integers}, i.e.   $n_\nu\in \theta_\nu+{\bf Z}$, where the
 $\theta_\nu$ are arbitrary constants.}
 \be \label{eq:rho}
 \rho_\ell = \prod_{\nu}
 \left(\frac{p_\nu+\ell}{p_\nu-\ell}\right)^{n_\nu}\    .
 \ee
 Thus we can impose for ${\bf C}$ the following type of discrete
 evolution
 \bse\label{eq:CC}\bea
 \wt{\bf C}\cdot (p-\tLd) &=& (p+\Ld)\cdot {\bf C}\   ,
 \label{eq:CCa} \\
 \wh{\bf C}\cdot (q-\tLd) &=& (q+\Ld)\cdot {\bf C}\   ,
 \label{eq:CCb} \eea\ese
 in which the shift ~${\bf C}\mapsto \wt{\bf C}$~  is the shift in the
 discrete variable $n$ associated with
 a (complex) parameter $p$ and the shift ~${\bf C}\mapsto
 \wh{\bf C}$~  is the shift in the discrete variable $m$ associated
 with the (complex) parameter $q$. In principle we can select an
 arbitrary set of lattice parameters $p_\nu$, each associated with
 its own lattice shift, and in view of the linearity
 of the corresponding equations for ${\bf C}$  we can impose all
 the discrete evolutions for all chosen values $p_\nu$ simultaneously.
 The true objects of interest, however, will obey nonlinear equations,
 and these objects are the following:
 \begin{itemize}
 \item The infinite matrix
 \be \label{eq:U}
 {\bf U} = {\bf C}\cdot \left( {\bf 1} + {\bf \Omega}\cdot {\bf
 C}\right)^{-1}\  ,
 \ee
 \item
 An infinite determinant
 \be\label{eq:tau}
 \tau\equiv \det\,_{\!\bf Z}\left( {\bf 1}+{\bf \Omega}\cdot
 {\bf C}\right)\  .
 \ee
 \end{itemize}
 To make sense of the latter we can use the expansion
 \[
 \det({\bf 1}+A)=1+\sum_i A_{ii} + \sum_{i<j} \left|
 \begin{array}{cc}A_{ii}&A_{ij}\\
 A_{ji}&A_{jj}\end{array}\right| + \dots \   ,
 \]
 and imposing that the integrations in (\ref{eq:C}) are such that the terms
 in the expansion which
 are all of the form
 \[ {\rm tr}_{\bf Z}\left( (\Omega\cdot {\bf C})^k\right)  \]
 truncates. The relevant quantities in terms of which one can derive
 closed-form nonlinear
 equations are actually the individual entries of the infinite matrix ${\bf
 U}$, which --by
 the way-- is symmetric, $\,^{t\!}{\bf U}={\bf U}$ as a consequence of the
 symmetry of ${\bf C}$.
 
 \subsection{Basic Relations}
 
 To derive the basic equations from this scheme, we mention the
 following relations that can be derived from the definitions (for
 details, cf. \cite{Carg}). For the $\tau$-function we have
 \be\label{eq:TAU}
 \frac{\wt{\tau}}{\tau} = \det\,_{\!\bf Z}\left( {\bf 1}+ (p-\tLd)^{-1}\cdot
 {\bf O}\cdot {\bf U}\right)
 = 1 + \left({\bf U}\cdot (p-\tLd)^{-1}\right)_{0,0}\   ,
 \ee
 whereas the infinite matrix $\bU$ obeys the matrix Riccati equation
 \be \label{eq:UU}
 \wt{\bf U}\cdot (p-\tLd) =
 (p+\Ld)\cdot {\bf U} - \wt{\bf U}\cdot {\bf O}\cdot {\bf U}\  . \
 \ee
 Eq. (\ref{eq:UU}) forms the starting point for the construction of
 integrable lattice
 equations. By combining different shifts associated with different
 parameters $p$
 one can actually derive all relevant discrete equations within the KdV family
 from this single relation. Eq. (\ref{eq:UU}) has its continuous counterpart
 as well
 which can be derived by a similar set of manipulations starting from
 (\ref{eq:CC}).
 
 Once we have selected a particular set of continuous or discrete
 variables we have yet another type of relations that we can
 consistently impose on the quantity ${\bf C}$, involving the operation
 generated by the matrices ${\bf I}$ and $\tII$.
 
 The dependence of the plane wave-factor $\rho_\ell$ on both the
 discrete variables $n_\nu$ as well as on the lattice parameters
 $p_\nu$ and the spectral variable $\ell$ is what determines the
 various equations for the relevant quantities. Clearly, from the
 definition (\ref{eq:rho}) these are given by
 \be\label{eq:Tnu}
 T_\nu\rho_\ell = \left(\frac{p_\nu+\ell}{p_\nu-\ell}\right) \rho_\ell
 \     ,
 \ee
 for the dependence of $\rho_\ell$ on the $n_\nu$, (the $T_\nu$ denoting
 the shift with respect to the variable $n_\nu$, i.e.
 $T_\nu\rho_\ell(n_\nu)=\rho_\ell(n_\nu+1)$). Furthermore, the
 dependence on the lattice variables is given by
 \be\label{eq:Tpnu}
 \frac{\partial}{\partial p_\nu}
 \rho_\ell = n_\nu \left( \frac{1}{p_\nu +\ell} -
 \frac{1}{p_\nu-\ell} \right) \rho_\ell\   ,
 \ee
 whereas the dependence on the spectral variable is given by
 \be
 \ell \frac{\partial}{\partial\ell}
 \rho_\ell = \sum_\nu n_\nu p_\nu \left( \frac{1}{p_\nu -\ell} -
 \frac{1}{p_\nu+\ell}
 \right) \rho_\ell\   .  \label{eq:Rb}
 \ee
 Note that from (\ref{eq:Tnu}) and (\ref{eq:Tpnu}) we immediately have
 that the operations of shift in $n_\nu$ and differentiation
 w.r.t. $p_\nu$ commute:
 \[ T_\nu\frac{\partial}{\partial p_\nu} \rho_\ell =
 \frac{\partial}{\partial p_\nu}T_\nu \rho_\ell\   .  \]
 Keeping in mind the representation (\ref{eq:C}) for the infinite
 matrix  ${\bf C}$, we are led to the following choice of linear
 constraint:
 \bea\label{eq:CI}
 \left[\ell\rho_\ell{\bf c}_\ell \tcl \right]_{\partial\Gamma}
 &=& {\bf C}+ {\bf I}\cdot {\bf C} + {\bf C}\cdot \tII  \nn \\
 && ~ + \sum_\nu p_\nu n_\nu
 \left( {\bf C}\cdot\frac{1}{p_\nu-\tLd} -
 \frac{1}{p_\nu+\Ld}\cdot {\bf C}  \right)\  ,
 \eea
 reflecting the scaling invariance. The term on the left-hand side
 is the boundary term from the integrations over the curve $\Gamma$
 in the complex $\ell$-plane. We don't want to assume on this formal
 level too much concerning the analytic details of the functions
 involved within the integrations, and thus we will make the
 assumption that the boundary terms are primarily dominated by
 the behaviour of the integrants in the limits as $\ell$ tends
 either to zero or to infinity. Thus, formally, we assume that the
 boundary terms can be interpreted in the following way
 \[ \left[ f(\ell)\right]_{\partial\Gamma} =
 \left( \ld \lim_{\ell\rightarrow\infty} - \mu
 \lim_{\ell\rightarrow 0}\right) f(\ell)\   , \]
 for any ``regular'' function $f$.
 It will turn out that with this assumption these terms will lead
 {\it precisely} to the correct behaviour accounting for the
 appearence of the free parameters in the corresponding
 nonlinear equations as a consequence of the asymptotic
 behaviour of the factors $\rho_\ell$ as $\ell\rightarrow 0$ and
 as $\ell\rightarrow\infty$, namely
 \[ \rho_\ell\  \ \rightarrow \  \ \left\{ \ \begin{array}{lcr}
 1+2\ell\sum_\nu (n_\nu/p_\nu) + {\cal O}(\ell^2) &\  \ {\rm as}
 \   \ & \ell\rightarrow 0\\
 (-1)^{\sum_\nu n_\nu}\left[1 + 2\ell^{-1}\sum_\nu p_\nu n_\nu +
 {\cal O}(\ell^{-2}) \right] &\  \ {\rm as}\   \ & \ell\rightarrow \infty
 \end{array}\right.   \   . \]
 
 In addition to (\ref{eq:CI}) we have the
 dependence on the lattice parameters, in terms of which ${\bf C}$
 obeys the following linear differential equations
 \be\label{eq:dpC}
 \frac{\partial}{\partial p_\nu}{\bf C} =
 n_\nu\left( \frac{1}{p_\nu+\Ld}\cdot{\bf C} -
 {\bf C}\cdot\frac{1}{p_\nu-\tLd} \right)\   .
 \ee
 as a consequence of which the discrete equation will be compatible
 with continuous equations in terms of the lattice parameters.
 To derive the corresponding equation for ${\bf U}$ we need to multiply
 (\ref{eq:CI}) from the left by $ ({\bf 1}+{\bf \Omega}\cdot
 {\bf C})^{-1}={\bf 1}-{\bf \Omega}\cdot {\bf U}$ and use
 eq. (\ref{eq:Omega}) to derive the following nonlinear counterpart
 of (\ref{eq:CI})
 \bea\label{eq:UI}
 \left[\ell\rho_\ell^{-1} {\bf u}_\ell \tul \right]_{\partial\Gamma}
 &=& {\bf U}+{\bf I} \cdot {\bf U}+{\bf U} \cdot\tII  \nn \\
 && ~+ \sum_\nu
 n_\nu p_\nu  \left( {\bf U}\cdot\frac{1}{p_\nu-\tLd} -
 \frac{1}{p_\nu+\Ld}\cdot{\bf U} -
 {\bf U} \cdot \frac{1}{p_\nu -\tLd}\cdot {\bf O}\cdot
 \frac{1}{p_\nu+\Ld}\cdot {\bf U}\right) \   , \nn \\
 \eea
 as well as the nonlinear counterpart of (\ref{eq:dpC})
 \be\label{eq:dpU}
 \frac{\partial}{\partial p_\nu}{\bf U} =
 n_\nu\left( \frac{1}{p_\nu+\Ld}\cdot{\bf U} -
 {\bf U}\cdot\frac{1}{p_\nu-\tLd} +
 {\bf U}\cdot\frac{1}{p_\nu-\tLd}\cdot {\bf O}\cdot
 \frac{1}{p_\nu+\Ld}\cdot{\bf U}
 \right)\   .
 \ee
 In the boundary terms of eq. (\ref{eq:UI}) we have introduced the
 infinite vector
 \be\label{eq:ul}
 {\bf u}_\ell = \rho_\ell \left( {\bf 1} - {\bf U}\cdot\Omega\right)
 \cdot {\bf c}_\ell\   .
 \ee
 It will turn out that the vector ${\bf u}_\ell$ will be the
 starting point for the construction of the monodromy problems
 for the resulting lattice systems. In fact, components of this
 vector will form the eigenfunctions of the corresponding Lax pairs,
 and the variable $\ell$ will play the role of the spectral
 parameter.
 
 Now that we have the framework in place from which the similarity
 reduction on the lattice can be derived, we can
 exploit the relations derived earlier to get closed-form discrete
 equations, as well as their similarity reduction.
 The KdV family is characterized by the symmetry
 \[ {\bf C}=\,^{t\!}{\bf C}\  \ \Rightarrow\  \
 {\bf U}=\,^{t\!}{\bf U}\    , \]
 leading to an additional set of algebraic relations of the form
 \be \label{eq:cUU}
 {\bf U}\cdot (-\tLd)^j = \Ld^j\cdot {\bf U} -
 {\bf U}\cdot {\bf O}_j\cdot {\bf U}\    \  ,\    \ j\ \ {\rm even}\  .  \ee
 This condition on the infinite matrix system will be specific
 to the KdV class of lattice equations.
 In the next sections we will employ the various relations for the
 infinite matrix ${\bf U}$ to derive closed-form equations for some of
 its entries leading to on the one hand
 integrable lattice equations as on the other hand their similarity
 constraints in closed form.

 \subsection{Lattice Equations}

 Let us now introduce several objects in terms of which one can derive
 from the basic system (\ref{eq:UU}) closed-form equations, namely
 \bse\label{eq:id}\be\label{eq:idU}
 u\equiv {\bf U}_{0,0}\    \ ,\    \
 s_{\ar,\bb}\equiv \left( \frac{1}{\ar + \Ld} \cdot
 {\bf U} \cdot \frac{1}{\bb + \tLd}\right)_{0,0}  \  ,
 \ee
 as well as
 \be\label{eq:vs}
 v_\ar\equiv 1-\left( \frac{1}{\ar + \Ld} \cdot {\bf U}\right)_{0,0}
 \     \ , \     \ s_\ar\equiv\ar-\left( \frac{1}{\ar+\Ld}\cdot{\bf U}
 \cdot\tLd\right)_{0,0}\   .
 \ee \ese
 In (\ref{eq:id}) $\ar$ and $\bb$ are arbitrary parameters which we
 can choose at our convenience. The KdV case is distinguished by the fact
 that the variable $s_{\ar,\bb}$ is symmetric in the parameters $\ar$
 and $\bb$, i.e. $s_{\ar,\bb}=s_{\bb,\ar}$.
 It is obvious from the definitions
 that $v_\ar$ can be obtained from $s_{\ar,\bb}$
 in the limit ~$\bb\rightarrow \infty$~, and that $u$ is recovered
 in the simultaneous limit ~$\ar,\bb\rightarrow\infty$~. However,
 we shall be interested as well in cases where either $\ar=0$ or
 $\bb=0$ or both.
 
 {}From the basic equations (\ref{eq:UU}) we can now derive the
 following set of relations for the objects defined in
 (\ref{eq:id}):
 \bse \label{eq:relst} \bea
 \widetilde{s}_\ar &=& (p+u)\widetilde{v}_\ar - (p-\ar) v_\ar\   \\
 s_\bb &=& (p+\bb) \widetilde{v}_\bb - (p-\widetilde{u}) v_\bb\   ,
 \eea \ese
 as well as
 \be
 1 - (p+\bb)\widetilde{s}_{\ar,\bb} + (p-\ar)s_{\ar,\bb}\,=\,
 \widetilde{v}_\ar v_\bb \  , \label{eq:svw}
 \ee
 in which the tilde denotes the shift $T_\nu$ with respect to a
 discrete variable $n_\nu=n$ associated with the lattice parameter
 $p_\nu=p$. Introducing a second discrete variable $n_\nu=m$ associated
 with the lattice parameter $p_\nu=q$, we have two different
 transformations on the lattice,namely
 \[ \rho_k \mapsto \widetilde{\rho}_k\   \ ,\    \
  \rho_k \mapsto \widehat{\rho}_k\   ,\]
 which can be combined in order to obtained closed-form equations.
 Since the dependence on the two variables $n$ and $m$ enter in
 ${\bf C}$ via the $\rho_\ell$ of (\ref{eq:rho}),
 it is clear that ${\bf C}$ depends linearly on these
 variables implying the permutability of the lattice shifts. This
 allows us by combining two lattice shifts to derive closed-from
 partial difference equations. In fact, all objects
 (\ref{eq:idU}) and (\ref{eq:vs}) are then interpreted as functions
 of the lattice sites $(n,m)$, and the shifts indicated by
 ~$\wt{\cdot}$~ and ~$\wh{\cdot}$~ represent the translations
 \[ (n,m)\mapsto (n+1,m)\   \ ,\   \ (n,m)\mapsto (n,m+1)\   . \]
 Thus, for the objects (\ref{eq:idU}) and (\ref{eq:vs})
 we derive from the various relations (\ref{eq:relst}) and
 (\ref{eq:svw}) nonlinear partial difference equations leading
 to the various lattice equations within the KdV family.
 
 The most general lattice equation that can be derived directly
 from (\ref{eq:svw}) is
 \be
 \frac{ 1-(p+\bb)\wt{s}_{\ar,\bb}+(p-\ar)s_{\ar,\bb}}{
 1-(q+\bb)\wh{s}_{\ar,\bb}+(q-\ar)s_{\ar,\bb}}\,
 =\, \frac{ 1-(q+\ar)\wh{\wt{s}}_{\ar,\bb}+(q-\bb)\wt{s}_{\ar,\bb}}{
 1-(p+\ar)\wh{\wt{s}}_{\ar,\bb}+(p-\bb)\wh{s}_{\ar,\bb} }\   ,
 \label{eq:s}
 \ee
 cf. \cite{NQC}, which for fixed values of $\ar$ and $\bb$
 is an integrable  partial difference equation for $s_{\ar,\bb}$ with
 as independence variables the lattice sites $(n,m)$. The $p$ and $q$
 can be interpreted as the corresponding lattice
 parameters. It must be pointed out that the independent discrete
 variables $n$ and $m$, are attached to the lattice parameters $p$
 respectively $q$, so that fixing the latter means choosing certain
 lattice directions of an in principle multidimensional grid. As is
 clear from (\ref{eq:rho}), we are absolutely
 free to introduce additional lattice directions each
 carrying its own lattice parameter. In terms of each pair of these lattice
 directions we have an equation of the form (\ref{eq:s}) for one
 and the same object $s_{\ar,\bb}$, and all these equations can be shown to
 be compatible. From that perspective the equation (\ref{eq:s})
 actually represents a {\it parameter-family} of compatible
 equations, which is the precise analogue of the {\it hierarchy} of
 nonlinear evolution equations in the case of the continuous KdV
 systems.
 
 Eq. (\ref{eq:s}) has several specialisations
 according to the choice of the latter parameters. The equations
 satisfied by the variables $u$ and $v_\ar$ follow from the limits
 ~$\ar\rightarrow\infty$, $\bb\rightarrow\infty$~, respectively
 ~$\ar$ arbitrary, $\bb\rightarrow\infty$~. Between these variables
 there are relations of Miura type which can be derived from
 (\ref{eq:relst}), eliminating the intermediate variable $s_\ar$
 by using both translation on the lattice, namely
 \bse \label{eq:miura} \bea
 p - q +\wh{u} -\wt{u} &=& (p-\ar) \frac{\wh{v}_\ar}{\wh{\wt{v}}_\ar}
 -(q-\ar) \frac{\wt{v}_\ar}{\wh{\wt{v}}_\ar}  \nn  \\
 &=& (p+\bb)\frac{\wt{v}_\bb}{v_\bb}-
 (q+\bb)\frac{\wh{v}_\bb}{v_\bb} \    , \label{eq:miuraa}  \\
 p + q + u -\wh{\wt{u}} &=&(p-\ar)\frac{v_\ar}{\wt{v}_\ar}+(q+\ar)
 \frac{\wh{\wt{v}}_\ar}{\wt{v}_\ar}   \nn \\
 &=& (p+\bb)\frac{\wh{\wt{v}}_\bb}{\wh{v}_\bb}+(q-\bb)
 \frac{v_\bb}{\wh{v}_\bb}\   .  \label{eq:miurab}
 \eea\ese
 
 Within the lattice KdV family we will concentrate on the three
 main equations, namely the ones for for $u$, $v_0$ and $z$, where
 $z$ is related to $s_{0,0}$ by
 \be\label{eq:z} z=s_{0,0}-\frac{n}{p}-\frac{m}{q},   \ee
 because it is for these variables that closed-form similarity
 constraints can be derived. The lattice equations satisfied by
 these objects can be easily deducted from the relations given above
 or directly from (\ref{eq:s}) by the appropriate limits. The
 explicit forms will be given in section 3.1.
 
 {}From the point of view of considering (\ref{eq:s}) and its
 atavars as parameter-families of compatible lattice equations,
 it makes sense also to consider the
 continuous equations expressing the dependence of the variables
 $u$, $v$ and $s$ on the lattice parameters. The resulting differential
 equations are by construction compatible with the lattice equations and
 integrable by themselves. They follow from the basic equations
 \bse\label{eq:conteq}\bea
 \frac{\partial u}{\partial p} &=& n\left( 1-v_pv_{-p}
 \right) \   , \label{eq:conteqa}  \\
 \frac{\partial v_\ar}{\partial p} &=& -n\left( \frac{v_p-v_\ar}{p-\ar}
 +s_{\ar,-p}v_p \right) \   , \label{eq:conteqb}  \\
 \frac{\partial s_{\ar,\bb}}{\partial p} &=& n\left(
 \frac{s_{\ar,\bb}-s_{p,\bb}}{p-\ar}
 - \frac{s_{\ar,\bb}-s_{\ar,-p}}{p+\bb}-s_{\ar,-p}s_{p,\bb} \right) \
    , \label{eq:conteqc}
 \eea\ese
 leading in particular to the differential-difference equations
 for the variables with $\ar,\bb=0,\infty$, namely
 \bse\label{eq:conteqs}\bea
 \frac{\partial u}{\partial p} &=& n\left( 1-\frac{2p}{2p+\undertilde{u}
 -\wt{u}} \right) \   , \label{eq:conteqsa}  \\
 \frac{\partial}{\partial p} \log\,v_0&=& -\frac{n}{p}
 \frac{\wt{v}_0-{\hypotilde 0 v_0}}{\wt{v}_0+{\hypotilde 0 v_0}}
 \   , \label{eq:conteqsb}  \\
 \frac{\partial z}{\partial p} &=& -\frac{2n}{p}
 \frac{(\wt{z}-z)(z-{\hypotilde 0 z})}{\wt{z}-{\hypotilde 0 z}}
 \   . \label{eq:conteqsc}
 \eea\ese
 Obviously, similar equations hold in terms of the parameter $q$.
 
 Finally, we mention the bilinear equations that are obeyed
 by the $\tau$-function, that was formally  defined by
 the representation (\ref{eq:tau}) from which one can deduce
 the following identifications of the variables
 $v_p$ and $v_q$ in terms of the $\tau$-function
 \be\label{eq:vtau}
 v_p=\frac{{\hypotilde 0 \tau}}{\tau}\     \ ,\      \
 v_q=\frac{{\hypohat 0 \tau}}{\tau} \   .
 \ee
 (The undertilde ${\hypotilde 0 \cdot}$ and underhat ${\hypohat 0 \cdot}$
 denote the lattice translations in the reverse directions
 to the shift $\wt{\cdot}$ respectively $\wh{\cdot}$~).
 In fact, we can associate with every value of the parameter $\ar$, instead
 of $p$ or $q$, a variable $v_\ar$  by similar expressions as (\ref{eq:vtau})
 in terms of lattice translations associated with that value.
 In particular, with $\ar=0$ we obtain the identification
 \be\label{eq:v0}
 v_0 = \frac{\tau_-}{\tau_+}\   ,
 \ee
 in which
 \be\label{eq:tau+-}
 \tau_\pm\equiv
 \det\,_{\!\bf Z}\left( {\bf 1} \pm \Omega\cdot{\bf C} \right)\  .
 \ee
 The $\tau_-$ can be viewed as the $\tau$-function obtained from
 $\tau=\tau_+$, i.e. (\ref{eq:tau}), by applying a lattice shift
 associated with
 parameter value $\ar=0$. For both $\tau_\pm$ we have the following
 bilinear lattice equations
 \bse\label{eq:taueqs}\bea
 (p+q)\wh{\tau}\wt{\hypohat 0 \tau} &=& 2q\tau\wt{\tau}
 + (p-q){\hypohat 0  \tau} \wh{\wt{\tau}} \   , \label{eq:taueqsa} \\
 (p+q)\wt{\tau} \wh{\hypotilde 0 \tau} &=& 2p\tau\wh{\tau}
 - (p-q){\hypotilde 0 \tau} \wh{\wt{\tau}} \   , \label{eq:taueqsb}
 \eea\ese
 which, together with the identifications (\ref{eq:vtau}), are
 easily obtained from the equation for $v_p$, namely
 \be\label{eq:veqs}
 (p+q)\frac{\wh{v}_p}{\wt{v}_p} + (p-q) \frac{v_p}{\wh{\wt{v}}_p}
 = 2p\   ,
 \ee
 respectively from a similar equation for $v_q$. The bilinear
 equations (\ref{eq:taueqsa}) and (\ref{eq:taueqsb}) are not
 independent  but there is a consistency that follows from
 the following diagram
 \vspace{.5cm}
 
 \setlength{\unitlength}{.5mm}
 \begin{picture}(180,60)(-60,0)

 \put(120,0){\circle*{3}}
 \put(150,0){\makebox(0,0){$\circ$}}
 \put(180,0){\makebox(0,0){$\otimes$}}
 \put(120,30){\circle*{3}}
 \put(150,30){\circle*{3}}
 \put(180,30){\makebox(0,0){$\times$}}
 \put(120,60){\circle*{3}}
 \put(150,60){\circle*{3}}
 \put(180,60){\circle*{3}}
 
 \put(120,0){\line(1,0){60}}
 \put(120,0){\line(0,1){60}}
 \put(150,0){\line(0,1){30}}
 \put(180,0){\line(0,1){60}}
 \put(120,30){\line(1,0){60}}
 \put(120,30){\line(0,1){30}}
 \put(150,30){\line(0,1){30}}
 \put(180,30){\line(0,1){30}}
 \put(120,60){\line(1,0){60}}
 \put(0,0){\circle*{3}}
 \put(0,30){\circle*{3}}
 \put(30,30){\circle*{3}}
 \put(0,60){\circle*{3}}
 \put(30,60){\circle*{3}}
 \put(60,60){\circle*{3}}
 
 
 \end{picture}
 \vspace{.2cm}
 
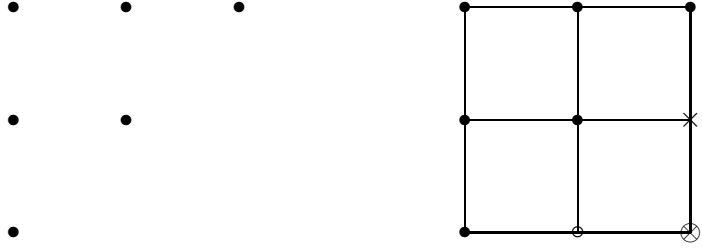
 \centerline{{\bf Figure 1:} {\it Consistency of the bilinear
 equations}.}
 \vspace{.3cm}
 
 \noindent In fact, taking initial data at the vertices of the
 first diagram in Fig. 1, we can calculate the remaining vertices in
 the rectangle vertically by using (\ref{eq:taueqsa}) and horizontally by
 using (\ref{eq:taueqsb}). Then the value of $\tau$ at the vertex
 indicated by ($\otimes$) at the right-bottom corner can be
 calculated in two independent ways, but it
 is easily checked by explicit calculation that this does not
 lead to an inconsistency: we obtain the same value irrespective
 of the way in which we calculate that value.
 
 Mixed equations between $\tau_+$ and $\tau_-$ can be derived from
 formulae like:
 \be\label{eq:mixv}
 \frac{pv+q\wh{\wt{v}}}{\wt{v}}=(p+q)\frac{\wh{\wt{v}}_p}{\wt{v}_p}
 \      \ ,\      \
 \frac{p\wt{v}-q\wh{v}}{v}=(p-q)\frac{\wt{v}_p}{\wh{\wt{v}}_p}
 \   ,
 \ee
 which follow from (\ref{eq:miuraa}) and (\ref{eq:miurab}) for
 different values of $\ar$ and $\bb$. Eqs. (\ref{eq:mixv}) lead to
 the following bilinear relations
 \bse\label{eq:mixtau}  \bea
 (p+q)\wh{\tau}_\pm\wt{\tau}_\mp&=&p\wh{\wt{\tau}}_\pm\tau_\mp +
 q\wh{\wt{\tau}}_\mp\tau_\pm\   .    \label{eq:mixtaua} \\
 (p-q)\wh{\wt{\tau}}_\pm\tau_\mp&=&p\wh{\tau}_\pm\wt{\tau}_\mp -
 q\wh{\tau}_\mp\wt{\tau}_\pm\   .    \label{eq:mixtaub}
 \eea\ese
 
 For $\tau$ we have also differential relations in terms of the
 lattice parameters. In fact, from the definition (\ref{eq:tau})
 one can easily deduce that
 \be\label{eq:taup}
 \ddp \log\tau = n s_{p,-p}\  ,
 \ee
 and from (\ref{eq:svw}) for $\ar=-p$, $\bb=p$, together with
 (\ref{eq:vtau}) we then have the bilinear equations of the form
 \bse\label{eq:taupeq}\bea
 \wt{\wt{\tau}}\ut{\tau}&=&\wt{\tau}\tau +\frac{2p}{n}\wt{\tau}
 \frac{\pl\tau}{\pl p} - \frac{2p}{n+1}\tau\frac{\pl\wt{\tau}}{\pl p}
 \   , \label{eq:taupeqa} \\
 \wh{\wt{\tau}}\ut{\tau}&=&\wh{\tau}\tau +\frac{p+q}{n}\left(
 \wh{\tau} \frac{\pl\tau}{\pl p}-\tau\frac{\pl\wh{\tau}}{\pl p}
 \right)\   , \label{eq:taupeqb}
 \eea\ese
 which will provide us with bilinear equations of
 differential-difference type.

 \subsection{Similarity Constraints}
 
 To obtain closed-form constraints for  single objects we need
 the cases that the parameters $\ar$ and $\bb$ go either to zero or
 infinity (otherwise we will get constraints containing derivatives
 with respect to the variables $\ar$ and $\bb$). So, we will focus
 primarily on similarity constraints for the objects: $u$, $v_0$ and
 $s_{0,0}$, but we need some other objects along the way as well.
 The similarity constraints for $u$, $v_0$ and $z$ are obtained from
 \bse\label{eq:constrs} \bea
 &&2\ld(-1)^{n+m}(np+mq-u)= u+np( v_pv_{-p}-1)+mq( v_qv_{-q} -1)\   ,
  \label{eq:SCu}  \\
 &&\left(\mu-\ld (-1)^{n+m}\right) v_0 =
 n\left[ (1+ps_{0,-p})v_p-v_0\right] +
 m\left[ (1+qs_{0,-q})v_q-v_0\right] \   , \nn \\  \label{eq:SCv}  \\
 &&-(2\mu+1) z=\frac{n}{p} (1+ps_{0,-p})(1-ps_{p,0})
 +\frac{m}{q} (1+qs_{0,-q})(1-qs_{q,0})\   ,  \label{eq:SCz}
 \eea\ese
 with (\ref{eq:z}), and in which
 we need to use the relations
 \bse\label{eq:st}\bea
 1+ps_{0,-p} &=& \wt{v}_0v_{-p} \   , \label{eq:sta} \\
 1-p\wt{s}_{p,0} &=& \wt{v}_pv_0 \   , \label{eq:stb} \\
 -(p-\wt{u})v_{-p} &=& s_{-p}\   , \label{eq:stc}
 \eea\ese
 and various other expressions that follow directly from
 (\ref{eq:relst}), (\ref{eq:svw}), such as
 \be\label{eq:vv} \wt{v}_pv_{-p}=1\    \ ,\     \
 2p+\ut{u}-\wt{u}=p\frac{\wt{v}_0+{\hypotilde 0 v}_0}{v_0}
 =2p\frac{\wt{v}_p}{v_p}\   ,
 \ee
 (and similar equations with $\wt{\cdot}$ replaced by $\wh{\cdot}$ and
 $p$ replaced by $q$). In the next section we will collect the explicit
 forms of the resulting equations.
 
 \paragraph{Remark 2.1:} In the similarity constraints 
(\ref{eq:constrs})
 we have restricted ourselves to the situation where there are only
 two discrete variables $n$ and $m$ associated with lattice paramters
 $p$ and $q$ respectively. However, we are absolutely free to extend the
 equations to include other lattice variables associated with additional
 directions (each carrying its own lattice parameter). For each of
 such lattice direction we have to {\it add on} a term of the same form
 as the ones present on the right-hand side of eqs. (\ref{eq:SCu}),
 (\ref{eq:SCv}) and (\ref{eq:SCz}). In the physics of
 two-dimensional quantum gravity this
 has its analogue in the famous {\it string equation}, where it
 has occurred in the literature on random matrix models, cf. e.g.
 \cite{FIK,Per}. Effectively, the procedure of increasing the number
 of variables in the similarity constraints amounts to building what
 is the discrete analogue of the {\it Painlev\'e hierarchy}, which
 in the continuum leads to increasingly higher-order equations after
 the elimination of redundant variables. This point of view on
 discrete Painlev\'e hierarchies is interesting, but to develop it
 furher is beyond the scope of the present paper.

 \section{The Lattice Painlev\'e System}
 \setcounter{equation}{0}
 
 On the basis of the structure exposed in the previous section,
 we will present in this section the closed-form expressions
 for the similarity reductions of the members of the lattice
 KdV family, which comprises of the three classes of equations:
 those for the variables $u$, $v$ and $z$ which will be identified
 with respectively the lattice potential KdV equation, the
 lattice potential MKdV equation and the lattice Schwarzian
 KdV (SKdV) equation. On the continuous level these equations are
 given respectively by:
 \bse\label{eq:Conteqs}\bea
 u_t&=&u_{xxx}+3 u^2_x\   , \label{eq:KdV}  \\
 v_t&=&v_{xxx}-3\frac{v_xv_{xx}}{v}\   , \label{eq:MKdV}  \\
 \frac{z_t}{z_x} &=& S(z)\equiv \frac{z_{xxx}}{z_x} - \frac{3}{2}
 \frac{z_{xx}^2}{z_x^2}\   . \label{eq:SKdV}
 \eea\ese
 Eq. (\ref{eq:MKdV}) is the potential MKdV equation, i.e.
 the variable $\bar{v}=(\log v)_x$ obeys the MKdV equation. Eq.
 (\ref{eq:SKdV}), (where $S(z)$ denotes the Schwarzian derivative
 of $z$), is the SKdV equation which is
 invariant under M\"obius-transformations:
 \be\label{eq:Mobius}
 z\  \ \mapsto\  \ \frac{az+b}{cz+d}\    \ ,\   \ ad-bc\neq 0\  .
 \ee
 As is well-known, these equations are interrelated via Miura
 transformations and effectively form the Miura chain:
 \vspace{.1cm}
 
 \begin{center}
 {\sf SKdV}\hspace{.3cm} $\longrightarrow$ \hspace{.3cm}
 {\sf MKdV}\hspace{.3cm} $\longrightarrow$ \hspace{.3cm}
 {\sf KdV}\hspace{.3cm}
 \end{center}
 
 \noindent
 In fact, in the first step the connections is given by the
 Hopf-Cole transformation
 \be\label{eq:HC}
 z_x=v^2\    \ \Rightarrow\    \
 \bar{v}= \frac{1}{2}\,\frac{z_{xx}}{z_x}\     \ \Leftrightarrow \     \
 \bar{v}_x-\bar{v}^2=\frac{1}{2} \frac{z_t}{z_x}\   ,
 \ee
 whereas in the second step we have the Miura transformation
 \be\label{eq:Miura}
 u_x=\frac{1}{2} S(z)=\frac{v_{xx}}{v}-2\frac{v_x^2}{v^2}=
 \bar{v}_x-\bar{v}^2\   .
 \ee
 
 The much-heralded conjecture of \cite{ARS} postulating the
 connection between similarity reductions of integrable nonlinear
 evolution equations and ODE's having the Painlev\'e property has been
 the motivation in \cite{NP} for pursuing a similar connection on the
 discrete level. Rather than trying to find directly a similarity
 variable (which on the discrete level is rather awkward) it was
 found to be more convenient to formulate the reduction in terms
 of constraints on the solution of the lattice equations. To
 illustrate the idea by means of the continuous equations
 (\ref{eq:Conteqs}), it is clear that the similarity reduction
 under scaling symmetry of the SKdV equation, which, as was noted
 in \cite{Dorf}, turns out to be of the form
 \be\label{eq:Simil}
 z(x,t)= t^{\bar{\mu}} Z(\xi)\    \ ,\    \ \xi=xt^{-1/3}\   ,
 \ee
 where $\bar{\mu}=\frac{1}{3}(1+2\mu)$ is an arbitrary constant (the
 parameter $\mu$ we introduce for later purpose), can be cast in
 the form of a linear nonautonomous constraint, namely
 \be \label{eq:2.9} 3\bar{\mu} z = xz_x + 3t z_t\   . \ee
 Imposing (\ref{eq:2.9}) on the solutions of (\ref{eq:SKdV}) and
 eliminating the $t$-derivatives we obtain the following third-order ODE
 \be\label{eq:SPII}
 S(z)= \frac{\bar{\mu}}{t} \frac{z}{z_x} -\frac{x}{3t}\  .
 \ee
 Eq. (\ref{eq:SPII}), in which $t$ is now merely playing the role of
 a parameter of the equation, was referred to in \cite{Dorf} as
 ``Schwarzian PII'' (SPII), because of its relation to the Painlev\'e
 II (PII) equation. In fact, via the Hopf-Cole transformation
 (\ref{eq:HC}) we obtain from  (\ref{eq:SPII}) the Painlev\'e II
 (PII) equation
 \be\label{eq:PII}
 \bar{v}_{xx}= 2 \bar{v}^3-\frac{x}{3t}\bar{v}+\frac{\mu}{3t}\   ,
 \ee
 which by itself can be obtained also from a linear constraint,
 namely
 \be\label{eq:vconstr}
 \mu v=xv_x+3tv_t\   \ \Rightarrow\   \
 \bar{v}+x\bar{v}_x+3t\bar{v}_t=0\   ,
 \ee
 which represents the well-known reduction of the MKdV to PII, cf.
 \cite{AS}. To complete the picture, we mention that the similarity
 constraint for the potential KdV is given by
 \be\label{eq:uconstr}
 u+xu_x+3tu_t= f(t)\   ,
 \ee
 with the freedom to have an arbitrary function $f(t)$ on the
 right-hand side.
 Thus, we obtain for ~$U=u_x+\frac{x}{6t}$~ Painlev\'e 34 (P34), i.e.
 the equation no. 34 in Painlev\'e's list, \cite{Pain,Gambier}, namely
 \be \label{eq:P34}
 UU_{xx}-\frac{1}{2} U_x^2+2U^3-\frac{x}{3t}U^2+
 \frac{\bar{\mu}^2}{8t^2}=0
 \   .
 \ee
 {}From the Miura transformation (\ref{eq:Miura}) between the MKdV and
 KdV equation one easily obtains the Miura and inverse Miura
 transformations relating PII to P34, \cite{FokAb}, namely
 \be\label{eq:invMiura}
 \bar{v}=\frac{\frac{\bar{\mu}}{2t}-U_x}{2U}\     \ \Leftrightarrow
 U=\bar{v}_x-\bar{v}^2+\frac{x}{6t}\   .
 \ee
 
 In the next subsection we will give the precise discrete analogues
 of these formulae and demonstrate that the entire Miura scheme,
 as well as its similarity reduction, survives on the discrete level.
 The main difference between the continuous situation and the
 discrete one lies in the fact that all similarity constraints
 become {\it nonlinear} on the discrete level, and therefore are
 not easily solved by extracting an explicit similarity variable.
 The actual forms of
 the constraints are highly nontrivial and follow from the infinite
 matrix structure of the previous section. Nonetheless, as we will
 show explicitely, all the interrelations between the various
 discrete equations and similarity constraints will exist on the
 discrete level and can, therefore, be checked independently
 from the equations themselves without reference to the structure
 developed in section 2.
 
 \subsection{Explicit Forms}
 
 The first example of a lattice equation we consider is the
 {\it cross-ratio equation} \cite{KDV}
 \be\label{eq:CR}
 \frac{(z_{n,m}-z_{n+1,m})(z_{n,m+1}-z_{n+1,m+1})}{(z_{n,m}-z_{n,m+1})
 (z_{n+1,m}-z_{n+1,m+1})} = \frac{q^2}{p^2}\    ,
 \ee
 which is obtained from (\ref{eq:s}) by taking $\ar=\bb=0$.
 This equation expresses the condition that under the map ~$(n,m)\mapsto
 z_{n,m}$~ the canonical cross-ratio of four adjacent points on the
 lattice is constant, which was interpreted recently as a notion of
 {\it discrete holomorphicity}, \cite{Bobenko} in connection
 with its invariance under (\ref{eq:Mobius}).
 Since it is well-known that the cross-ratio forms the discrete
 analogue of the Schwarzian derivative, eq.   (\ref{eq:CR})
 is the natural discrete version of the Schwarzian KdV equation
 (\ref{eq:SKdV}).
 
 In the discrete case the similarity constraint for the lattice
 Schwarzian KdV (\ref{eq:CR}) takes on the form
 \be \label{eq:skdvconstr}
 (1+2\mu) z_{n,m}=2n\frac{(z_{n+1,m}-z_{n,m})(z_{n,m}-z_{n-1,m})}{ z_{n+1,m}
 - z_{n-1,m} }
 +  2m \frac{ (z_{n,m+1} - z_{n,m})(z_{n,m} - z_{n,m-1})}{ z_{n,m+1} -
 z_{n,m-1} }\   ,
 \ee
 which was first given in \cite{Dorf}. As was
 pointed out in \cite{Bobenko2}, this nonautonomous nonlinear constraint
 coincides precisely with the condition that the map
 ~$(n,m)\mapsto z_{n,m}$~ can be identified with the discrete
 analogue of the {\it power map} ~$\zeta\mapsto \zeta^a$~ as a
 discrete conformal map in the sense of \cite{Bobenko}. This is clear
 by comparison with the continuum situation, where it resides
 in the fact that the time-dependent prefactor in the solution of
 the similarity constraint, i.e. (\ref{eq:Simil}), is given exactly
 by the power map $t\mapsto t^{\bar{\mu}}$ in the complex $t$-plane.
 
 Eq. (\ref{eq:CR}) can be trivially resolved by identifying the
 differences in the variables $z$ in terms of products of variables
 $v$ by means of the following relations
 \bse\label{eq:dHC}\bea
 p(z_{n,m}-z_{n+1,m})&=&v_{n+1,m}v_{n,m}\   , \\
 q(z_{n,m}-z_{n,m+1})&=&v_{n,m+1}v_{n,m}\   ,
 \eea\ese
 which in turn by eliminating the $z$ variables in an additive way
 leads to the following nonlinear equation for the new variable $v=v_{n,m}$,
 (which coincides with the variable $v_0$  in section 2),
 \be \label{eq:dmkdv}
 pv_{n,m+1}v_{n,m} + qv_{n+1,m+1}v_{n,m+1}
 = qv_{n+1,m}v_{n,m} + pv_{n+1,m+1}v_{n+1,m} \   ,
 \ee
 which is the lattice version of the potential MKdV equation (\ref{eq:MKdV}).
 The similarity constraint (\ref{eq:skdvconstr}) can then
 be used to obtain a similarity constraint in terms of $v$. In fact,
 taking the difference of (\ref{eq:skdvconstr}) in the
 $n$-variable and using (\ref{eq:dHC}) we obtain
 \begin{eqnarray*}
 \mu+\frac{1}{2} &=& (n+1)\frac{v_{n+2,m}}{v_{n+2,m}+v_{n,m}} -
 n\frac{v_{n-1,m}}{v_{n+1,m}+v_{n-1,m}}  \\
 && + m\frac{p}{q}\left(
 \frac{v_{n+1,m+1}v_{n+1,m-1}}{v_{n,m}(v_{n+1,m+1}+v_{n+1,m-1})}
 -\frac{v_{n,m+1}v_{n,m-1}}{v_{n+1,m}(v_{n,m+1}+v_{n,m-1})}
 \right) \   ,
 \end{eqnarray*}
 which by using next (\ref{eq:dmkdv}) to transform the last two terms
 can be cast into the form
 \begin{eqnarray*}
 \mu&=& \frac{1}{2}(n+1)\frac{v_{n+2,m}-v_{n,m}}{v_{n+2,m}+v_{n,m}} +
 \frac{1}{2}n\frac{v_{n,m}-v_{n-1,m}}{v_{n+1,m}+v_{n-1,m}}  \\
 && + m
 \frac{v_{n+1,m+1}v_{n,m+1}-v_{n+1,m-1}v_{n,m-1}}{(v_{n+1,m+1}+v_{n+1,m-1})
 (v_{n,m+1}+v_{n,m-1})}\   ,
 \end{eqnarray*}
 where the right-hand side is actually the sum of two terms plus
 their $n$-shifted counterparts. Thus, after performing one
 discrete ``integration'' we obtain
 the following similarity constraint in terms of $v_{n,m}$
 \be\label{eq:mkdvconstr}
 n\frac{v_{n+1,m}-v_{n-1,m}}{v_{n+1,m}+v_{n-1,m}} +
 m\frac{v_{n,m+1}-v_{n,m-1}}{v_{n,m+1}+v_{n,m-1}}  =
 \mu -\ld (-1)^{n+m} \   ,
 \ee
 where $\ld$ appears as an integration constant. (Obviously, the whole
 argument goes equally through with $m$-shifts as well as with
 the $n$-shifts). The alternating term arising from a discrete
 integration is typical for the discrete situation
 and adds effectively an extra parameter to the equation.
 The resulting similarity constraint for $v_{n,m}$ is exactly
 the one that one obtains from the formalism of section 2, namely
 eq. (\ref{eq:SCv}).
 
 It is on this level of the equation for $v_{n,m}$ that one can
 actually eliminate the
 dependence on one or the other of the discrete variables $n$ or $m$
 in the discrete case. In fact, concentrating on the $n$-dependence,
 we define as dependent variables
 \be\label{eq:x-y}  x_n = x_n(m)\equiv
 \frac{v_{n,m}}{v_{n+1,m+1}}\      \ ,
 y_n = x_n(m)\equiv \frac{v_{n+1,m}}{v_{n,m+1}}\      \ ,   \ee
 which through the lattice MKdV equation (\ref{eq:dmkdv})
 are related via
 \be\label{eq:xy} x_n=\frac{y_n-r}{1-ry_n} \    \ \Leftrightarrow \    \
 y_n=\frac{x_n+r}{1+rx_n} \   ,  \ee
 with $r\equiv q/p$, as well as the combinations
 \be\label{eq:a-b}
 a_n=a_n(m)\equiv\frac{v_{n+1,m}-v_{n-1,m}}{v_{n+1,m}+v_{n-1,m}}
 \     \ ,\     \
 b_n=b_n(m)\equiv\frac{v_{n,m+1}-v_{n,m-1}}{v_{n,m+1}+v_{n,m-1}}
 \   . \ee
 In order to eliminate the $m$--shifts from the equations
 we need to express $b_n$ in terms of $x_n$ or (equivalently)
 $y_n$. To achieve that we add to
 both sides of eq. (\ref{eq:dmkdv}) the corresponding terms
 of its $m$--backshifted counterpart which leads to the following
 relation
 \be\label{eq:B}  (x_n+r)b_{n+1}+x_n = (y_n-r)b_n + y_n \   . \ee
 Noting further that $a_n$ can be easily expressed in terms
 of $x_n$ (or $y_n$) through the relation ~$y_n/x_{n-1}=
 v_{n+1}/v_{n-1}$~, leading to
 \be\label{eq:axy}  a_n=\frac{y_n-x_{n-1}}{y_n+x_{n-1}}\   , \ee
 we can now use the similarity constraint (\ref{eq:mkdvconstr})
 to eliminate $b_n$ from (\ref{eq:B}) to obtain finally
 a closed-form ordinary difference equation in terms of either
 the variable $x_n$ or $y_n$ (or both), which reads
 \bea \label{eq:mpii}
 && (n+1)(r+x_n)(1+rx_n)
 \frac{x_{n+1}-x_n+r(1-x_nx_{n+1})}{x_{n+1}+x_n+r(1+x_nx_{n+1})} \nn \\
 && ~~~~~~~~~~~~~~ -n(1-r^2)x_n
 \frac{x_n-x_{n-1}+r(1-x_nx_{n-1})}{x_n+x_{n-1}+r(1+x_nx_{n-1})}= \nn \\
 && ~~~ = \mu r(1+2rx_n+x_n^2)+\ld(-1)^{n+m}(r+2x_n+rx_n^2)
 -mr(1-x_n^2)\   ,
 \eea
 in which $r=q/p$, $m$, $\mu$ and $\ld$  are parameters of the equation.
 In subsection 3.2 we show that (\ref{eq:mpii}) can, in fact, be
 regarded to be a discrete analogue of both the PIII as well as PV, and that in
 special limits of the parameters it contains
 both the discrete Painlev\'e II equation (dPII),
 eq. (\ref{eq:dPII}), as well as its alternate version (alt-dPII),
 eq. (\ref{eq:alt-dPII}).
 We conclude that eq. (\ref{eq:mpii}) seems to
 unify the various branches of discrete Painlev\'e equations.
 To obtain the equation in the form (\ref{eq:MPII}) we proceed
 slightly differently, namely by redefining the variables according
 to the odd/even lattice sites and setting
 \be\label{eq:odd-even}
 X_{2n}\equiv x_{2n}\     \ ,\    \ X_{2n+1}\equiv
 -\frac{1}{y_{2n+1}}\  .
 \ee
 Using then the relations (\ref{eq:B}) and (\ref{eq:axy}), as well
 as (\ref{eq:xy}) for the even and odd sites, eq. (\ref{eq:MPII})
 follow after a straightforward calculation.

 Arriving now at the equations for the variable $u$, we have the
 following partial difference equation
 \be
 \left( p-q+u_{n,m+1}-u_{n+1,m}\right)(p+q+u_{n,m}-
 u_{n+1,m+1} ) = p^2 -q^2\  ,
 \label{eq:dkdv}
 \ee
 which is the lattice (potential) KdV equation, which can easily
 derived from the relations in section 2. Eq. (\ref{eq:dkdv})
 resolves trivially the lattice MKdV equation, i.e. the lattice
 equation for $v_{n,m}$ in the form (\ref{eq:dMKdV}),
 via the Miura relations
 \bse \label{eq:2.8} \bea
 p - q + u_{n,m+1} -u_{n+1,m} &=& p\frac{v_{n,m+1}}{v_{n+1,m+1}}  -
 q\frac{v_{n+1,m}}{v_{n+1,m+1}} \    ,  \label{eq:2.8a} \\
 p + q + u_{n,m} -u_{n+1,m+1} &=& p\frac{v_{n,m}}{v_{n+1,m}}  +
 q\frac{v_{n+1,m+1}}{v_{n+1,m}} \    .   \label{eq:2.8b}
 \eea \ese
 The similarity reduction of the KdV is obtained from the
 constraint
 \be\label{eq:kdvconstr}
 (\ld (-1)^{n+m} + \frac{1}{2})(u_{n,m} -np-mq)+
 \frac{np^2}{2p+u_{n-1,m}-u_{n+1,m}} +
 \frac{mq^2}{2q+u_{n,m-1}-u_{n,m+1}} = 0\   ,
 \ee
 noting that the free parameter, $\ld$, now enters purely in the
 alternating term. Thus, in a sense the similarity constraint for
 $v_{n,m}$ interpolates between two extreme cases: the equation for
 $z_{n,m}$ in which only the parameter $\mu$ enters, and the equation
 for $u_{n,m}$ in which only the parameter $\ld$ is present.
 We have already shown that the constraints (\ref{eq:skdvconstr}) and
 (\ref{eq:mkdvconstr}) are related through the lattice Hopf-Cole
 transformations (\ref{eq:dHC}). It can also be shown that the
 constraints (\ref{eq:mkdvconstr}) and (\ref{eq:kdvconstr}) are
 related, namely via the lattice Miura transformations
 (\ref{eq:2.8}). In fact, using relations of the type (\ref{eq:vv})
 we can rewrite (\ref{eq:kdvconstr}) in the form
 \be\label{eq:kdvc}
 (\ld (-1)^{n+m} + \frac{1}{2})(u_{n,m} -np-mq)+
 \frac{npv_{n,m}}{v_{n+1,m}+v_{n-1,m}} +
 \frac{mqv_{n,m}}{v_{n,m+1}+v_{n,m-1}} = 0\   ,
 \ee
 and then by subtracting from the $n$-shifted counterpart
 of (\ref{eq:kdvconstr}) its $m$-shifted counterpart and using
 (\ref{eq:2.8a}) together with the lattice MKdV equation
 (\ref{eq:dmkdv}), the relation
 \[
 \frac{pv_{n,m}+qv_{n+1,m+1}}{v_{n+1,m+1}+v_{n+1,m-1}} =
 q+\frac{pv_{n+1,m}-qv_{n,m+1}}{v_{n,m+1}+v_{n,m-1}}
 \]
 and a similar formula with $p$ and $q$ and their respective
 shifts interchanged, we can derive the following
 \[
 pv_{n+1,m}\left( C_{n+1,m}-C_{n,m}\right) =
 qv_{n,m+1}\left( C_{n,m+1}-C_{n,m}\right) \  ,
 \]
 where $C_{n,m}$ stands for the constraint, i.e.
 ~$C_{n,m}\equiv na_n+mb_n-\mu+\ld(-1)^{n+m}$~. Alternatively,
 we can subtract from (\ref{eq:kdvc}) its $n$-- {\it and} $m$--
 shifted counterpart and use (\ref{eq:2.8b}) to derive
 \[
 pv_{n,m}\left( C_{n+1,m}-C_{n,m}\right) =
 qv_{n+1,m+1}\left( C_{n+1,m+1}-C_{n+1,m}\right) \  ,
 \]
 and obviously there is a similar relation with $p$ and $q$ and
 the $n$-- and $m$--shifts interchanged. Clearly, if $C_{n,m}$
 is identically zero all these relations are trivially satisfied
 demonstrating the consistency of the constraint (\ref{eq:kdvconstr})
 with the constraint (\ref{eq:mkdvconstr}).
 
 \paragraph{Remark 3.1:} As is evident from the comparison with the
 continuous situation (see previous subsection) the system
 consisting of (\ref{eq:dkdv}) and (\ref{eq:kdvconstr}) can be
 viewed as forming a discretisation of the
 discrete P34, (\ref{eq:P34}). The elimination of the $m$-shifted
 objects in favour of $n$-shifted ones yields an algebraic
 equation, and we will not give it here.
 
 Finally, we mention the bilinear forms for the lattice KdV
 family. They are given in terms of coupled system for two
 $\tau$-functions, $\tau$ and $\sg$, which are identified with
 $\tau_+$ respectively $\tau_-$ given in (\ref{eq:tau+-}),
 i.e. $\tau_+=\tau_{n,m}$ and $\tau_-=\sg_{n,m}$. For each
 these we have the following bilinear lattice equations
 \bse\label{eq:Taueqs}\bea
 (p+q)\tau_{n,m+1}\tau_{n+1,m-1} &=& 2q\tau_{n,m}\tau_{n+1,m}
 + (p-q)\tau_{n,m-1}\tau_{n+1,m+1} \   , \label{eq:Taueqsa} \\
 (p+q)\tau_{n+1,m}\tau_{n-1,m+1} &=& 2p\tau_{n,m}\tau_{n,m+1}
 - (p-q)\tau_{n-1,m}\tau_{n+1,m+1} \   , \label{eq:Taueqsb}
 \eea\ese
 and exactly the same equations for $\sg_{n,m}$. The coupling
 between $\tau$ and $\sg$ takes place via the additional relations
 \bse\label{eq:addrel}\bea
 \sg_{n+1,m}\tau_{n-1,m}+\sg_{n-1,m}\tau_{n+1,m}&=&
 2\sg_{n,m}\tau_{n,m}\  ,  \label{eq:addrela} \\
 \sg_{n,m+1}\tau_{n,m-1}+\sg_{n,m-1}\tau_{n,m+1}&=&
 2\sg_{n,m}\tau_{n,m}\   ,  \label{eq:addrelb}
 \eea\ese
 which follow from (\ref{eq:Taueqs}) by taking instead of
 the $n$-- or $m$--shift a shift in a third direction associated
 with a lattice parameter equal to zero, which amounts to a
 B\"acklund transformation $\tau_{n,m}\mapsto \sg_{n,m}$.
 More generally, we obtain from (\ref{eq:mixtau}) the
 coupled system
 \bse\label{eq:addrel2}\bea
 p\sg_{n,m}\tau_{n+1,m+1}+q\sg_{n+1,m+1}\tau_{n,m}&=&
 (p+q)\sg_{n+1,m}\tau_{n,m+1}\  , \label{eq:addrel2a} \\
 p\sg_{n+1,m}\tau_{n,m+1}-q\sg_{n,m+1}\tau_{n+1,m}&=&
 (p-q)\sg_{n,m}\tau_{n+1,m+1}\   ,  \label{eq:addrel2b}
 \eea\ese
 as well as two similar equations with $\sg$ and $\tau$ interchanged.
 Finally, the similarity constraints, which for the $\tau$-functions
 are given by the coupled system of bilinear equations:
 \bse\label{eq:tausim} \bea
 \left( n+m+\mu-\ld(-1)^{n+m}\right) \sg_{n,m}\tau_{n,m}
 &=& n\sg_{n+1,m}\tau_{n-1,m} + m \sg_{n,m+1}\tau_{n,m-1}\   , \nn \\
 \label{eq:tausima} \\
 \left( n+m-\mu+\ld(-1)^{n+m}\right) \sg_{n,m}\tau_{n,m}
 &=& n\sg_{n-1,m}\tau_{n+1,m} + m \sg_{n,m-1}\tau_{n,m+1}\   , \nn \\
 \label{eq:tausimb}
 \eea\ese
 which in conjunction with the lattice bilinear equations
 (\ref{eq:taueqs}) and the relations (\ref{eq:addrel2}) yield the
 bilinearisation of the lattice Painlev\'e system, and thereby of
 the GDP (\ref{eq:MPII}). The bilinear form (\ref{eq:tausim}) of
 the similarity constraint of \cite{NP} were first given in
 \cite{SRG}.
 We return to bilinear forms in section 5 within the context of
 B\"acklund- and Schlesinger transformations for the general lattice
 Painlev\'e system. We will there extend the bilinear formulation
 treating both independent variables $n$, $m$ and free parameters
 $\ld$, $\mu$ on the same footing, and demonstrate the important
 phenomenon of {\it self-duality} for the system in this extended
 bilinear form.
 
 \paragraph{Remark 3.2:} It is remarkable that the similarity constraint
 for the variable $z$, i.e. (\ref{eq:skdvconstr}) can be reformulated
 in various ways. In fact, since we have shown that together with
 the lattice equation (\ref{eq:CR}) one can actually {\it derive} the
 similarity constraint for $v$, leading to (\ref{eq:mkdvconstr}),
 via the Miura relations (\ref{eq:dHC}) which resolve the equation
 (\ref{eq:CR}), we can then retransform the constraint for $v$, again
 by using the Miura relations, into a seemingly different similarity
 constraint for $z$, namely
 \be \label{eq:skdvconstr2}
 \mu-\ld(-1)^{n+m}=n\frac{z_{n+1,m}+z_{n-1,m}-2z_{n,m}}
 { z_{n+1,m} - z_{n-1,m} }
 +m\frac{z_{n,m+1}+z_{n,m-1}-2z_{n,m}}{ z_{n,m+1} - z_{n,m-1} }\   ,
 \ee
 and pushing this even further, we can then combine
 (\ref{eq:skdvconstr2}) with the original constraint
 (\ref{eq:skdvconstr}) to obtain the following third form of the
 similarity constraint
 \be \label{eq:skdvconstr3}
 \left( 1+2\ld(-1)^{n+m}\right) z_{n,m} =
 2n\frac{z_{n,m}^2-z_{n+1,m}z_{n-1,m}}{ z_{n+1,m}-z_{n-1,m} }
 +2m\frac{z_{n,m}^2-z_{n,m+1}z_{n,m-1}}{z_{n,m+1}-z_{n,m-1} }\   .
 \ee
 Obviously, the three constraints (\ref{eq:skdvconstr}),
 (\ref{eq:skdvconstr2}) and (\ref{eq:skdvconstr3}) are not
 independent since we have indicated how to derive one from
 the other. However, it remains an amusing  fact that these three
 forms, each involving the Painlev\'e parameters $\mu$ and
 $\ld$ in a different way, exist and are equally valid.

 \subsection{Continuum Limits}

 The general discrete Painlev\'e equation (\ref{eq:mpii}) is a quite
 rich equation in that for special values of the parameters it
 contains previously known discrete Painlev\'e equations, namely
 the discrete analogues of the PII equation\footnote{
 For this reason we have often in private referred to (\ref{eq:mpii}) as
 the {\sl master} d-P$_{\rm II}$ equation.}.  In fact, taking the limit
 $r\rightarrow\infty$, $m/r\rightarrow\xi$, it is easy to see that
 (\ref{eq:mpii}) reduces to
 \be\label{eq:altdpii}
 \frac{n+1}{x_nx_{n+1}+1}+\frac{n}{x_nx_{n-1}+1} =
 n+\frac{1}{2}+\mu+\frac{1}{2}\xi\left(x_n-\frac{1}{x_n}\right)\   ,
 \ee
 which is the the alternate dPII equation (\ref{eq:alt-dPII}).
 
 The limit to
 the dPII equation (\ref{eq:dPII}) is a bit more subtle, since we need
 to work in an oblique direction. In fact, to obtain dPII from
 (\ref{eq:mpii}) we write $r=1+\dd$ and take the limit $\dd\rightarrow
 0$, whilst taking $n=n^\prime-m$, where $n,m\rightarrow\infty$
 such that $n^\prime$ is fixed (in the limit) and $\dd m\rightarrow
 \eta$ finite. In that limit we have also $(1-x_n)/(1+x_n)\rightarrow
 a_{n+1}$ and (\ref{eq:mpii}) reduces to the following equation:
 \be\label{eq:limit}
 (n^\prime+1)(1-x_n^2)-(\mu+(-\ld)^{n^\prime})(1+x_n)^2 =
 2\eta x_n\left(\frac{1-x_{n+1}}{1+x_{n+1}}+\frac{1-x_{n-1}}
 {1+x_{n-1}}\right)\   ,
 \ee
 where we consider $x_n$ to be a function of $n^\prime$ rather
 than $n$. In terms of $a_n$, (omitting from now on the primes
 on the $n^\prime$ variable), this equation reads
 \be\label{eq:dpii}
 \frac{1}{2}\eta(a_{n+1}+a_{n-1})=\frac{-\mu+\ld(-1)^n+na_n}{1-a_n^2}\  ,
 \ee
 which is the asymmetric dPII equation (\ref{eq:dPII}). In fact, this
 limit was
 the case that was considered originally in the paper \cite{NP}.

 However, by itself (\ref{eq:mpii})
 is richer than just d-P$_{\rm II}$, a fact that will be
 made clearer in the next section which deals with its Schlesinger
 transformations. In this subsection, we shall examine this equation
 at the simplest possible level: what are the continuous Painlev\'e
 equations contained in (\ref{eq:mpii})?
 Let us start with the symmetric form of (\ref{eq:mpii}) which we
 write as:
 \bea  \label{eq:MPIIa}
 && \zeta_{n+1}(x_n+r)(rx_n+1){x_{n+1}-x_n+r(1-x_nx_{n+1})\over
 x_{n+1}+x_n+r(1+x_nx_{n+1})}-\zeta_n(1-r^2)x_n{x_n-x_{n-1}
 +r(1-x_nx_{n-1})\over x_{n-1}+x_n+r(1+x_nx_{n-1})}  \nn \\
 && ~~~~=\mu r(1+2rx_n+x_n^2)-\eta_m(1-x_n^2)
 \eea
 Here $\zeta_n$ and $\eta_m$ are the explicit discrete variable
 related linearly to $n$ respectively $m$, i.e. in general form
 $\zeta_n=\dd n+\zeta_0$, $\eta_m=\dd m+\eta_0$.
 The introduction of this scaling of $n$ and $m$ is not unavoidable,
 but it makes the continuous limit procedure more straightforward.
 Next, we introduce the following substitutions:
 $x_n=w$, $t=n\epsilon$, $\dd=\epsilon$, $\zeta_0=\eta_0=0$,
 $\mu=\epsilon b$, $\zeta_m=\epsilon a$,  $r=\epsilon$, we obtain at the
 limit $\epsilon\to 0$ the following equation for $w$:
 \be\label{eq:PIII}
 w''={w'^2\over w}-{w'\over t}+{w^3}-{2a+2b+1\over t}w^2-{2b-2a-1\over
 t}-{1\over w}  \   ,
 \ee
 i.e. P$_{\rm III}$ in canonical form.
 
 We turn now to the asymmetric form of (\ref{eq:MPII}) which we write as:
 \bea \label{eq:MPIIb}
 &&\zeta_{n+1}(x_n+r)(rx_n+1){x_{n+1}-x_n+r(1-x_nx_{n+1})\over
 x_{n+1}+x_n+r(1+x_nx_{n+1})}-
 \zeta_n(1-r^2)x_n{x_n-x_{n-1}+r(1-x_nx_{n-1})\over
 x_{n-1}+x_n+r(1+x_nx_{n-1})}  \nn \\
 && ~~~=\mu
 r(1+2rx_n+x_n^2)+\nu(-1)^{n+m}(r+2x_n+rx_n^2)-\eta_m(1-x_n^2)\   .
 \eea
 Next, we introduce the following substitutions:
 $x_{n}=-1+\epsilon^4 w$ for even $n$, $x_n=-1+\epsilon^4 t/w+\epsilon^6
 u$ for odd $n$,
 where $t=\epsilon(n-m)$, $\dd=\epsilon^4$,
 $\zeta_0=\eta_0=1$, $\mu=\epsilon^4 (a-b)/2$,
 $\nu=-\epsilon^4 (a+b)/2$, $r=1+\epsilon^5$, we
 obtain at the
 limit $\epsilon\to 0$
 two equations containing $u$.
 Eliminating $u$ from these equations we get for $w$:
 \be\label{eq:piii}
 w''={w'^2\over w}-{w'\over t}+{w^3\over t^2}+{b w^2\over t^2}+{a\over
 t}-{1\over w}\   , \ee
 i.e. P$_{\rm III}$ although in a slightly non-canonical form.
 One may remark at
 this point that this continuous limit was obtained for $t=\epsilon(n-m)$
 and thus in effect, losing
 one of the parameters of the equation (namely $m$). It is not
 clear whether
 this is an absolute necessity for the continuous limit to exist or not.
 In the latter case another limit, to some ``higher'' discrete Painlev\'e
 equation might exist. We shall show in the
 following sections how one can obtain a P$_{\rm V}$ from (\ref{eq:MPII})
 and its
 Schlesinger transforms.
 
 \subsection{Connection with PVI}
 
 We now show that the similarity reduction also implies
 a reduction on the level of the continuous variables which
 in our case are the lattice parameters $p$ and $q$. In fact,
 from the relation (\ref{eq:conteqsb}) in conjunction with the
 relations given above we can now derive an ordinary differential
 equation for $y_n$ (and consequently $x_n$) with $p$ as
 independent variable. By construction this equation must
 be of Painlev\'e type, in fact related to PVI.
 Starting from (\ref{eq:conteqsb}) and its counterpart which can
 be written as
 \be\label{eq:dpv}
 -p\ddp \log v_{n,m}=na_n\   \ ,\   \ -q\ddq\log v_{n,m}=mb_n\   ,
 \ee
 and using ~$v_{n+1,m}/v_{n-1,m}=(1+a_n)/(1-a_n)$~ together with
 (\ref{eq:B}) we obtain
 \begin{eqnarray*} -p\ddp \log\left(\frac{1+a_n}{1-a_n}\right) &=&
 (n+1)a_{n+1} - (n-1)a_{n-1} = m(b_{n-1}-b_{n+1})  \\
 &=& (\mu-\ld(-1)^{n+m}-na_n)\left[ \frac{x_{n-1}+r}{y_{n-1}-r}
 - \frac{y_n-r}{x_n+r}\right]  \\
 && ~~ +  m \left[ \frac{x_{n-1}-y_{n-1}}{y_{n-1}-r}
 - \frac{y_n-x_n}{x_n+r}\right]\   ,
 \end{eqnarray*}
 which by using (\ref{eq:xy}) and the relation ~$x_{n-1}=y_n(1-a_n)/
 (1+a_n)$~ leads after some straightforward algebra to the following
 first-order equation for $a_n$ in terms of $a=a_n$ and $y=y_n$:
 \bse\be \label{eq:eqa}
 \frac{p}{r}(1-r^2) \frac{\pl a}{\pl p} =
 (-\mu+\ld(-1)^{n+m} + na-m)(1-a)y +
 (-\mu+\ld(-1)^{n+m} + na+m)\frac{1+a}{y}\   ,  \ee
 where we omitted now the label $n$ from all variables.
 On the other hand, by similar tricks we can derive a first-order
 differential equation for $y_n$, namely by proceeding as follows
 \begin{eqnarray*} -p\ddp \log y_n &=& (n+1)a_{n+1} - n\wh{a}_n
 = \mu+\ld(-1)^{n+m} -mb_{n+1}-n\wh{a}_n  \\
 &=& \mu+\ld(-1)^{n+m}+(\mu-\ld(-1)^{n+m}+na_n) \frac{y_n-r}{x_n+r} -
 m\frac{y_n-x_n}{x_n+r} \\
 && ~~ -  n \left[ \frac{1-y_nx_n}{y_n(x_n+1/r)}
 - \frac{y_n-r}{y_n(1+rx_n)}a_n\right]\   ,
 \end{eqnarray*}
 where $\wh{a}_n$ denotes as before the variable $a$ shifted in
 the $m$-direction and where we have used the counterpart of
 (\ref{eq:B}) in the $m$-direction, namely
 \[ \left(x_n+\frac{1}{r}\right)\wh{a}_n+x_n=\left(\frac{1}{y_n}
 -\frac{1}{r}\right) a_n+\frac{1}{y_n}\   .   \]
 Thus, we are led eventually to the the following expression
 for $a=a_n$ in terms of $y=y_n$:
 \be \label{eq:eqy}
 2na-\mu+\ld(-1)^{n+m} = \frac{(n+m)r(1-y^2)-(\mu+\ld(-1)^{n+m})
 (1-r^2)y-p(1-r^2)\frac{\pl y}{\pl p}}{(1-ry)(y-r)}\  ,
 \ee\ese
 which allows us to derive the following ODE from (\ref{eq:eqa})
 \bea\label{eq:PVI?}
 &&\frac{p}{r}(1-r^2)\ddp\left(
 \frac{(n+m)r(1-y^2)-(\mu+\ld(-1)^{n+m})
 (1-r^2)y-p(1-r^2)\frac{\pl y}{\pl p}}{(1-ry)(y-r)}\right) \nn \\
 && ~~~ = 2n(-\mu+\ld(-1)^{n+m} + na-m)(1-a)y +
 2n(-\mu+\ld(-1)^{n+m} + na+m)\frac{1+a}{y}\   , \nn  \\
 \eea
 where for $a$ we have to insert the expression (\ref{eq:eqy})
 in the right-hand side. The nonautonomous second-order ODE
 (\ref{eq:PVI?}) for $y$ as a function of the lattice parameter
 $p$ (which notably enters also via $r=q/p$), is
 directly related to PVI via
 \be\label{eq:PVIrel}
 w(t)=py(t)\    \ ,\     \ t=p^2\  ,
 \ee
 in terms of which, taking $q=1$ and abbreviating
 $\nu\equiv\ld(-1)^{n+m}$, we obtain precisely (\ref{eq:PVI})
 with the identifications
 \bea\label{eq:abcd}
 &&\ar=\frac{1}{8}(\mu-\nu+m-n)^2 ~~~~~\         \ ,\      \
 \bb=\frac{1}{8}(\mu-\nu-m+n)^2\   ,  \nn \\
 &&\gm=\frac{1}{8}(\mu+\nu-m-n-1)^2\     \ ,\      \
 \dd=\frac{1}{8}(\mu+\nu+m+n+1)^2\   .
 \eea
 
 The connection between the $y=y_n$ and PVI (\ref{eq:PVI})
 via (\ref{eq:PVIrel}) allows us to express the general solution
 of the GDP equation (\ref{eq:mpii}), or equivalently
 (\ref{eq:MPII}), in terms of PVI transcendents, namely by writing
 \be
 \label{eq:Soln} y_n=\frac{1}{p}P_{VI}(p^2;\ar,\bb,\gm,\dd;w_0,w_1)\  ,
 \ee
 and using the identifications (\ref{eq:abcd})
 and the relations (\ref{eq:xy}) and (\ref{eq:odd-even}) to obtain
 $x_n$ or $X_n$. The initial values $w_0=w(0)$ and $w_1=w^\prime(0)$
 can be identified with initial values for the discrete equation
 by using (\ref{eq:eqy}) together with (\ref{eq:axy}) and
 (\ref{eq:xy}). In fact, the latter corresponds to a Schlesinger
 transformation of the form ~$w^\prime=f(w,\ut{w},t)$~, ($w^\prime$
 denoting the derivative with respect to $t$ and $\ut{w}$ denoting
 the shift of the parameters $\ar$,$\bb$,$\gm$,$\dd$ according to
 $n\mapsto n-1$) which allows us to express for fixed $t$ the
 initial values of $w$ at $n=n_0$, (given by  parameters
 $\ar_0,\bb_0,\gm_0,\dd_0$ via (\ref{eq:abcd})), and at $n=n_0+1$
 (given by some different parameters  $\ar_1,\bb_1,\gm_1,\dd_1$)
 in terms of $w$ and $w^\prime$ at a fixed value of the parameters.
 Thus, if we assume the PVI transcendents to be known functions
 at each value of the parameters for given initial data $w_0$ and $w_1$,
 we have a two-parameter family solutions of the GDP. Of course,
 this starts from the assumption that all two-parameter solutions of
 the continuous PVI are known and fully under control as a class of
 nonlinear special functions, which is obviously not yet the case
 in this day and age. The initial value problem for PVI
 was considered in \cite{Jimbo} from the isomonodromic point of view.
 
 \paragraph{Remark 3.3:} It is well-known that the PVI equation
 (\ref{eq:PVI}) also admits a (Lie-point) symmetry group $G$ isomorphic
 to the symmetric group $S_4$ on four elements, \cite{Pain,Oka},
 cf. also \cite{IKSY}. This group $G$ is generated by three elements
 $g_i$, $i=1,2,3$, given by
 \bea\label{eq:group}
 g_1:&& ~~~ w\mapsto 1/w~~~~~~, ~~t\mapsto 1/t~~~~~~~ , ~~~~~
 (\ar,\bb,\gm,\dd)\mapsto (\bb,\ar,\gm,\dd)  \nn \\
 g_2:&& ~~~ w\mapsto 1-w~~~, ~~~t\mapsto 1-t~~~~~ , ~~~~~
 (\ar,\bb,\gm,\dd)\mapsto (\ar,\gm,\bb,\dd)  \\
 g_3:&& ~~~ w\mapsto \frac{1}{t}w~~~~~~, ~~~t\mapsto 1/t~~~~~~~
  , ~~~~~ (\ar,\bb,\gm,\dd)\mapsto (\ar,\bb,\dd,\gm).  \nn
 \eea
 This symmetry group $G$ was used in e.g. \cite{FY,FokAb}, cf.
 also \cite{GL}, to generate one-parameter families of special
 solutions of PVI.
 
 As can be seen from the identifications (\ref{eq:abcd}), on the
 discrete level these symmetries of PVI are very natural:
 the element $g_1$ corresponds to the interchange of the discrete
 variables $m$ and $n$; the composed map $g_1\circ g_3$
 corresponds to the transformation $\tau_{n,m}\leftrightarrow
 \sg_{n,m}$, $(\mu,\nu)\mapsto (-\mu,-\nu)$, which is the shift
 associated with lattice parameter $p_\nu=0$. The interpretation of
 the element $g_2$ is slightly more complicated: it can be inferred
 from the combination
 \[ g_3\circ g_2\circ g_1\circ g_2\circ g_3\circ g_2: ~~~ w\mapsto
 t\frac{w-1}{w-t}~~~~~,~~~t\mapsto t~~~~~,~~~~ (\ar,\bb,\gm,\dd)
 \mapsto (\dd,\gm,\bb,\ar)\  , \]
 which is equivalent to the replacement $y\mapsto -x$ and which
 corresponds to the replacement $n \rightarrow -n-1$. It is
 well-known, cf. e.g. \cite{JimboMiwa}, that the group of
 Schlesinger/B\"acklund transformations act on the solutions of
 PVI by discrete shifts which are the natural shifts in the
 variables $n,m$ as well as $\mu,\nu$. Thus, it seems that the entire
 symmetry group of PVI can be interpreted naturally in terms of
 operations on the lattice. We will make this point more poignantly
 in the next section where we will derive Schlesinger- and B\"acklund
 transformations for the GDP.
 
 \paragraph{Remark 3.4:}
 The bilinear form of PVI was found only recently, \cite{ORGT},
 even though $\tau$-function ``forms'' were presented earlier,
 cf. e.g. \cite{JimboMiwa,Okamo,HieKru}. We mention here that
 from the bilinear forms in subsection 3.1, namely eqs.
 (\ref{eq:Taueqs})-(\ref{eq:tausim}), together with the relations
 (\ref{eq:taupeq}), from which in particular one may deduce
 \bse\label{eq:PVIbil}\be  \label{eq:PVIa}
 \tau_{n,m}\sg_{n,m}=\tau_{n-1,m}\sg_{n+1,m}+\frac{p}{n}
 \left( \tau_{n,m}\frac{\partial\sg_{n,m}}{\partial p}
 - \sg_{n,m}\frac{\partial\tau_{n,m}}{\partial p} \right)\   ,
 \ee
 as well as
 \be\label{eq:PVIbilb}
 \tau_{n+1,m+1}\tau_{n-1,m}=\tau_{n,m+1}\tau_{n,m}+\frac{p+q}{n}
 \left( \tau_{n,m+1}\frac{\partial\tau_{n,m}}{\partial p}
 - \tau_{n,m}\frac{\partial\tau_{n,m+1}}{\partial p} \right)\   ,
 \ee \ese
 and an equation similar to (\ref{eq:PVIbilb}) for $\sg_{n,m}$.
 With the identifications, (taking as before $q=1$),
 \bea\label{eq:PVIid}
 &&w=p\frac{\sg_{n+1,m}\tau_{n,m+1}}{\sg_{n,m+1}\tau_{n+1,m}}
 =1-(1-p)\frac{\sg_{n,m}\tau_{n+1,m+1}}{\sg_{n,m+1}\tau_{n+1,m}}
 \   , \nn \\
 &&\frac{w-t}{1-t}= \frac{p}{1+p}\frac{\tau_{n,m}\sg_{n+1,m+1}}
 {\sg_{n,m+1}\tau_{n+1,m}}\   ,
 \eea
 the latter combination corresponding to the action of the element
 \[
 g_2\circ g_3\circ g_2: ~~~ w\mapsto \frac{w-t}{1-t}~~~,
 ~~t\mapsto \frac{t}{t-1}~~~~~
  , ~~~~~~ (\ar,\bb,\gm,\dd)\mapsto (\ar,\dd,\gm,\bb),
 \]
 of the symmetry group $G$ described above, eq. (\ref{eq:PVIbil})
 together with the total set of corresponding
 bilinear discrete equations for $\tau_{n,m}$ and $\sg_{n,m}$,
 (\ref{eq:Taueqs})-(\ref{eq:tausim}), provides the
 effective bilinear scheme for PVI. In
 \cite{ORGT} a bilinear system for PVI was obtained consisting of eight
 equations for eight different $\tau$-functions, which can be identified with
 the ones appearing (\ref{eq:PVIid}),
 including equations of second order in the continuous variable $t$.

 \section{Integrability Aspects}
 \setcounter{equation}{0}
 
 There are various aspects to the question regarding the integrability
 of the discrete systems presented in the previous section.
 Here we will discuss basically two of them: {\it i)} the issue of
 the consistency of the
 equations from the point of view of initial value problems, and
 {\it ii)} the isomonodromy  of the lattice Painlev\'e systems.
 Both aspects pertain to what is possibly the most intriguing
 question, namely: {\it to what extent do
 these discrete systems obey the Painlev\'e property?}. The
 Painlev\'e property being originally formulated for differential
 equations, \cite{Pain,Gambier}, it has been an important issue
 in recent years to formulate the analogous property for discrete
 systems, cf. \cite{GRP,CM,Krus}. In view of the fact that
 one of the main goals in the original studies by Painlev\'e was
 to use differential equations as a means of introducing new
 (transcendental) functions with appropriate uniformisability
 properties, cf. \cite{Pain2,Conte}, we believe that the issue of
 formulating the Painlev\'e property in its most general context
 (namely for difference equations as well as differential equations)
 constitutes one of the key steps in the general problem of
 developing the proper tools for the integration of difference
 equations.
 
 \subsection{Singularity Confinement and Consistency of the
 Evolution}
 
 One of the main breakthroughs in the search for the analogue of
 a Painlev\'e property for discrete systems has been
 the discovery in \cite{GRP} of the property of
 {\it singularity confinement} (SC). In coarse terms SC amounts to
 the phenomenon of non-propagation of singularities in integrable
 discrete systems. Whilst it is not at all clear whether SC is
 a sufficient condition for integrability, it can be argued
 that it is at least a necessary condition because it amounts
 to the well-posedness of the discrete evolution of the system.
 As an algorithmic tool it has proven to be very fruitful in
 predicting the integrability of many discrete systems, notably
 of most of the newly discovered discrete Painlev\'e equations,
 \cite{RGH,Capel}.
 
 To exemplify the singularity confinement mechanism we apply it
 to the situation given by the constrained lattice system derived
 in the previous sections. Here, we shall investigate the nonlinear
 system by itself without referring to the structures developed
 in section 2 from which it was derived. We need as a starting
 point the general form of the discrete system, allowing for
 arbitrary coefficients, conditions on which can then be found using
 the SC requirement. In the case of the nonlinear constraint  of the
 form (\ref{eq:mkdvconstr}), we may thus start from the ``bare''
 form of the constraint, \cite{NP},
 \be\label{eq:mkdvconstr0}
 n {v_{n+1,m}-v_{n-1,m}\over {v_{n+1,m}+v_{n-1,m}}}+m
 {{v_{n,m+1}-v_{n,m-1}\over
 {v_{n,m+1}+v_{n,m-1}}}}=0\   , \ee
 i.e. the form that leads to a discrete Painlev\'e equation
 without free parameters. This same
 form was analyzed in \cite{RGS} in the framework of the bilinear
 formalism. However, as we now know this is not the most general
 integrable form of the similarity constraint. Let us, therefore,
 investigate by what we can replace the r.h.s of
 (\ref{eq:mkdvconstr0}) to retain integrability, taking the point
 of view that the similarity constraint must be an integrable
 equation in its own right. Thus, taking instead of
 (\ref{eq:mkdvconstr0}) the form
 \be\label{eq:mkdvconstr1}
 n {v_{n+1,m}-v_{n-1,m}\over {v_{n+1,m}+v_{n-1,m}}}+m
 {{v_{n,m+1}-v_{n,m-1}\over
 {v_{n,m+1}+v_{n,m-1}}}}=A
 \ee
 where $A$ is in principle a function of $n$ and $m$. In order to
 apply the singularity confinement criterion we consider the
 possible singularities of ({\ref{eq:mkdvconstr1}):
 either $v$ becomes infinite or it  becomes 0. In both cases we
 require that the confinement is in one step, i.e. at the iteration
 immediately following the singular value we already obtain a finite
 value for $v$. This requirement leads to the
 following conditions for $A$:
 \begin{eqnarray*}
 &&(m^2-n^2)\left[m(A(n+1,m)-A(n-1,m))+n(A(n,m+1)-A(n,m-1))\right]\\
 &&+n[A(n+1,m)A(n-1,m)+1)(A(n,m+1)-A(n,m-1)]\\
 &&-m[A(n,m+1)A(n,m-1)+1)(A(n+1,m)-A(n-1,m)]\\
 &&+2nm[A(n+1,m)+A(n-1,m)-A(n,m+1)-A(n,m-1)]=0\   ,
 \end{eqnarray*}
 and
 \begin{eqnarray*}
 &&(m^2-n^2)\left[(A(n+1,m)-A(n-1,m))(A(n,m+1)-A(n,m-1)\right] \\
 &&+n[A(n+1,m)+A(n-1,m))(A(n,m+1)-A(n,m-1)]  \\
 &&-m[A(n+1,m)-A(n-1,m))(A(n,m+1)+A(n,m-1)]  \\
 &&+2nm[A(n+1,m)A(n-1,m)-A(n,m+1)A(n,m-1)]=0\   .
 \end{eqnarray*}
 The complete solution of this system is not of particular interest
 since we have already restricted the choice of the coefficients in
 the l.h.s. of (\ref{eq:mkdvconstr1})  to exactly $m$ and $n$ while
 they could a priori also have been any functions of $n$ and $m$.
 However, there exists a solution that  is really interesting in
 the case at hand, namely a constant $A$. Moreover, a look at
 (\ref{eq:mkdvconstr1}) suffices to convince oneself that this
 constant may have an even-odd dependence (since (\ref{eq:mkdvconstr1})
 involves points that are
 all shifted by an even number of steps of each other) We arrive thus
 to the precise form given in (\ref{eq:mkdvconstr}).
 
 Let us now turn to the issue of the consistency of lattice equation
 and similarity constraint. For the KdV family where the lattice
 equations involve vertices around elementary plaquettes, whereas the
 similarity constraints are given in terms of cross-like configurations,
 we have thus a system of equations that can be symbolically
 represented as
 \addtocounter{equation}{1}
 \vspace{.2cm}
 
 \setlength{\unitlength}{.5mm}
 \begin{picture}(150,60)(-50,0)
 \put(0,15){\circle*{3}}
 \put(0,45){\circle*{3}}
 \put(30,15){\circle*{3}}
 \put(30,45){\circle*{3}}
 \put(0,15){\line(1,0){30}}
 \put(0,15){\line(0,1){30}}
 \put(30,45){\line(-1,0){30}}
 \put(30,45){\line(0,-1){30}}
 
 \put(60,30){\makebox(0,0)[c]{$+$}}
 
 \put(90,30){\circle*{3}}
 \put(120,30){\circle*{3}}
 \put(150,30){\circle*{3}}
 \put(120,0){\circle*{3}}
 \put(120,60){\circle*{3}}
 \put(220,30){(\arabic{section}.\arabic{equation})}
 \put(90,30){\line(1,0){30}}
 \put(120,30){\line(1,0){30}}
 \put(120,30){\line(0,-1){30}}
 \put(120,30){\line(0,1){30}}
 \end{picture}
 \vspace{.3cm}
 
 \noindent
 As a starting point for the solution of initial value problems for
 systems of the form (4.3) 
 one may take initial data on the lattice according to the following
 configuration
 \addtocounter{equation}{1}
 \vspace{.2cm}
 
 \setlength{\unitlength}{.5mm}
 \begin{picture}(180,60)(-100,0)
 \put(0,45){\circle*{3}}
 \put(30,45){\circle*{3}}
 \put(30,15){\circle*{3}}
 \put(170,30){(\arabic{section}.\arabic{equation})\label{eq:initial}}
 \put(60,15){\circle*{3}}
 \put(0,45){\line(1,0){30}}
 \put(30,45){\line(0,-1){30}}
 \put(30,15){\line(1,0){30}}
 
 \end{picture}
 \vspace{.2cm}
 
 \noindent
 and use the lattice equation and
 constraint in conjunction to each other to fill the entire lattice.
 The issue of compatibility arises from the fact that at some points
 on the lattice the determination is not unique, and the
 uniqueness (i.e. single-valuedness of the solution around
 localised configurations) needs to be verified. In fact, this is a
 discrete manifestation of the Painlev\'e property, not quite formulated
 in the same way as the singularity confinement phenomenon that was put
 forward in \cite{GRP}. Our conjecture is that singularity confinement
 is a consequence of the single-valuedness in the above sense.
 
 The iteration of the system works as follows: starting with data
 given on a configuration of lattice points of the form  (4.4),
 one can proceed through the lattice by usinga
  the lattice equation together with the similarity constraint and
 calculate the values of the variable at each lattice point.
 In Figure 2 are indicated
 points that are calculated by means of the lattice equation by a cross
 (~$\times$~) and points that are calculated using the similarity
 constraint by an open circle (~$\circ$~).
 However, there are going to be points of possible conflict
 where we can calculate the value in two different ways. The first
 point where this happens is indicated in the figure
 by a cross within a circle, (~$\otimes$~). For the similarity
 reduction to be consistent, i.e. for the similarity constraint
 and the lattice equation to be {\it compatible}, the value at this
 point needs to be unambiguous i.e. both ways of calculating the
 value must give the same result.
 \vspace{.3cm}
 
 \setlength{\unitlength}{.5mm}
 \begin{picture}(90,90)(-90,0)
 \put(30,0){\circle{3}}
 \put(60,0){\makebox(0,0){$\times$}}
 \put(30,90){\circle{3}}
 \put(60,90){\makebox(0,0){$\times$}}
 \put(90,60){\circle{3}}
 \put(0,30){\makebox(0,0){$\times$}}
 \put(60,60){\makebox(0,0){$\times$}}
 \put(0,60){\circle*{3}}
 \put(30,30){\circle*{3}}
 \put(30,60){\circle*{3}}
 \put(60,30){\circle*{3}}
 \put(0,60){\line(1,0){30}}
 \put(30,60){\line(0,-1){30}}
 \put(30,30){\line(1,0){30}}
 \put(90,30){\makebox(0,0){$\otimes$}}
 \end{picture}
 \vspace{.2cm}
 
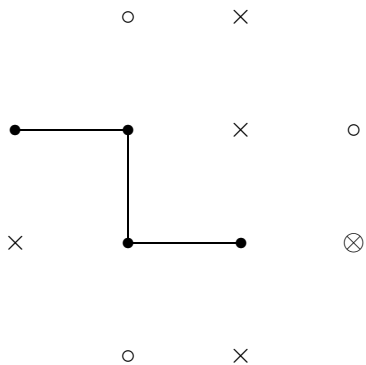
 \centerline{{\bf Figure 2:} {\it Consistency of the constrained
 lattice system}.}
 \vspace{.3cm}
 
 The linear system of partial difference equations
 \bse\label{eq:linsyst} \bea
 && n\left( f_{n+1,m}-f_{n-1,m}\right) +
 m\left( f_{n,m+1}-f_{n,m-1}\right) = \mu
 -\ld (-1)^{n+m} \   ,  \label{eq:linsysta} \\
 &&(p+q)\left( f_{n,m+1}-f_{n+1,m}\right) = (p-q)\left( f_{n+1,m+1} -
 f_{n,m} \right)\  ,  \label{eq:linsystb}
 \eea \ese
 is possibly the simplest system of the form (4.3) for
 which one can check the compatibility on the lattice by direct
 calculation. This system, in fact, is the linearised form of the
 systems of lattice KdV type as given in the previous section.
 Obvously, it is much harder to ascertain the compatibility
 for the full nonlinear system, and one needs to resort to algebraic
 manipulation programmes (we have used MAPLE and REDUCE ourselves)
 to establish the consistency in a direct way for arbitrary initial
 data. Thus, by direct calculation on the level of the equations,
 we may state the following:
 
 \paragraph{Proposition 4.1:} {\it The nonlinear systems of partial difference
 equations on the two-dimensional lattice associated with the KdV
 family, i.e. either (\ref{eq:CR}) together with (\ref{eq:skdvconstr}),
 or (\ref{eq:dmkdv}) together with (\ref{eq:mkdvconstr}), or
 (\ref{eq:dkdv}) together with (\ref{eq:kdvconstr}) are all compatible,
 i.e. arbitrary initial data given on configurations of the form (4.4)
 lead to a single-valued determination of values for
 the dependent variable on each vertex of the two-dimensional lattice.}
 \vspace{.2cm}
 
 In \cite{DIGP} it was argued that the compatibility of the
 system should be interpreted as a symmetry of the lattice
 equation. Thus, the lattice Painlev\'e reduction is a source of
 inspiration to a natural formulation of what could be a general
 definition of a symmetry of a partial difference equation.
 To elaborate a general theory of symmetries of discrete
 equations along these lines is a challenging problem for the future.

 \subsection{Isomonodromic Deformation Problems}
 
 In section 2 we have shown how
 to derive the lattice systems as well as their similarity constraints
 leading from the infinite matrix formalism. Subsequently,
 exploiting these lattice equations together with the constraints
 we have shown how to obtain the general discrete Painlev\'e
 equations, either in the form of (\ref{eq:MPII}) or as
 (\ref{eq:mpii}). In this section we will show that we can obtain
 from the infinite matrix structure also the associated
 linear problems which are the relevant isomonodromic deformation
 problems. In the spirit of the original works by Fuchs and
 Schlesinger \cite{Fuchs,Schles}  the integrability of the
 discrete Painlev\'e equations can thus be assessed in an independent
 way, namely by
 exploiting the monodromy problems and use them as a starting point for
 the asymptotic analysis to characterise the transcendental solutions,
 which for the continuous Painlev\'e equations has been an ongoing
 programme during the last two decades, cf. e.g.
 \cite{JMU}-\cite{FZ}. A similar programme for
 the discrete Painlev\'e equations has, in our opinion, to be
 developed as well, but this is beyond the scope of the
 present paper. We contend ourselves here by just providing the tools
 for the further analytic study of these nonlinear difference equations,
 which at present is still only in its initial stages. (Notably, to date
 the only discrete Painlev\'e equation that has been systematically
 investigated from an analytic point of view is dPI (\ref{eq:dPI}),
 cf. e.g. \cite{FIK,Bleher}).
 
 We will now show how isomonodromic deformation problems can be derived
 systematically on the basis of the infinite-matrix structure that we
 developed in section 2. Expanding that scheme further we can
 introduce a set of linear relations for infinite vectors which
 were already defined in (\ref{eq:ul}), namely
 \be\label{eq:uk}
 {\bf u}_k = \rho_k \left( {\bf 1} - {\bf U}\cdot\Omega\right)\cdot
 {\bf c}_k\   ,
 \ee
 having components $u_k^j$, ($j\in {\bf Z}$),
 with $\rho_k$ given in (\ref{eq:rho}) and ${\bf c}_k$ denoting,
 as before, the vector with entries ~$({\bf c}_k)_j=k^j$~.
 We will use the relations of section 2 to obtain linear systems of
 equations for ${\bf u}_k$ in terms of the infinite matrices defined
 there, and then extract from these infinite systems finite systems
 in terms of special entries, leading to concrete
 Lax pairs as well as monodromy problems.
 Recalling that the action
 of the operators $\Ld$ and ${\bf I}$ on ${\bf c}_k$ is given by
 \[ \Ld\cdot {\bf c}_k=k {\bf c}_k\     \ ,\      \
 {\bf I}\cdot {\bf c}_k= k\frac{d}{dk}{\bf c}_k\   , \]
 and making use of the relations (\ref{eq:interrels}) we can derive from the
 matrix Riccati equation (\ref{eq:UU}) for ${\bf U}$ the following
 {\it linear} relations for ${\bf u}_k$
 \bse\label{eq:linuk}\bea
 (p-k)\wt{\bf u}_k &=& \left( p+\Ld - \wt{\bf U}\cdot {\bf O}\right)\cdot
 {\bf u}_k\   , \\
 (p+k){\bf u}_k &=& \left( p-\Ld + {\bf U}\cdot {\bf O}\right)\cdot
 \wt{\bf u}_k\   .
 \eea\ese
 
 {}From eqs. (\ref{eq:linuk}) one can derive Lax pairs for
 the various lattice equations within the KdV family and the relation
 (\ref{eq:cUU}) can, in fact, be used to establish the gauge
 transformations between the various Lax pairs as will be shown
 below. Next we need to construct the similarity constraint for
 ${\bf u}_k$ which can be obtained in a similar way, now using
 the similarity constraint (\ref{eq:UI}) as a starting point.
 The result is the following
 \bea\label{eq:monouk}
 &&\left\{ \left[ \ell {\bf u}_\ell \tcl\cdot {\bf \Omega}\right]_{\partial
 \Gamma}  + {\bf I}\right\} \cdot {\bf u}_k = \nn \\
 && ~~~= k\frac{d}{dk}{\bf u}_k -
 \sum_\nu n_\nu p_\nu  \left[ \frac{1}{p_\nu-k} - \left(
 {\bf 1} + {\bf U} \cdot \frac{1}{p_\nu -\tLd}\cdot {\bf O}\right)
 \cdot \frac{1}{p_\nu+\Ld}\right] \cdot {\bf u}_k \   .
 \eea
 
 The next step is to derive concrete Lax pairs and monodromy problems
 from the infinite set of equations encoded in
 (\ref{eq:linuk})-(\ref{eq:monouk}).
 This can be achieved by singling out specific components, the choice
 of which will determine which equation within the lattice KdV family
 we are dealing with. In addition we need the relations pertaining to
 the dependence on the lattice parameters $p$ and $q$. These are given
 by the basic relations consisting of
 \be\label{eq:puk}
 \frac{\partial}{\partial p}{\bf u}_k = n\left(
 \frac{1}{p+\Ld} - \frac{1}{p-k} + {\bf U}\cdot\frac{1}{p-\tLd}\cdot
 {\bf O}\cdot\frac{1}{p+\Ld} \right) {\bf u}_k\   .
 \ee
 
 For the system consisting of the lattice MKdV (\ref{eq:dmkdv}) together
 with the corresponding similarity constraint (\ref{eq:mkdvconstr})
 we have the following monodromy problem \cite{Carg}. It consists of
 the linear differential equation in terms of the spectral parameter
 $k$, cf. \cite{NP},
 \bea
 k\frac{d}{dk}\psi_{n,m} &=& \left( \begin{array}{cc}
 -(1+\mu)&0\\ 0&\gm_{n+m}\end{array} \right) \psi_{n,m}
 +\frac{2nv_{n,m}}{v_{n+1,m}+v_{n-1,m}} \left(
 \begin{array}{cc} 0 & v_{n+1,m} \\  0& -p \end{array} \right)
 \psi_{n-1,m} \nn \\
 && ~~~~+\frac{2mv_{n,m}}{v_{n,m+1}+ v_{n,m-1}} \left(
 \begin{array}{cc} 0& v_{n,m+1}  \\ 0& -q \end{array} \right)
 \psi_{n,m-1}\   ,   \label{eq:MKdVmono}
 \eea
 where ~$\gm_n\equiv n+\ld(-1)^{n}$~, in the case
 of the lattice MKdV, together with the discrete Lax pair in the form
 of the form
 \bse\label{eq:2.3}  \bea
 \psi_{n+1,m} &=&  L_{n,m}(k^2) \psi_{n,m}\   , \label{eq:2.3a} \\
 \psi_{n,m+1} &=&  M_{n,m}(k^2) \psi_{n,m}\   , \label{eq:2.3b}
 \eea \ese
 obeying the discrete zero-curvature condition
 \be \label{eq:zerocurv}
 L_{n,m+1} M_{n,m} = M_{n+1,m} L_{n,m}\   .
 \ee
 The matrices $L$ and $M$ are given by
 \be \label{eq:MKdVlax}
 L_{n,m}(k^2)=\left(\begin{array}{cc}
 p& v_{n+1,m} \\ \frac{k^2}{v_{n,m}}&
 p\frac{v_{n+1,m}}{v_{n,m}}\end{array}
 \right) \     \ ,
 \     \
 M_{n,m}(k^2)=\left(\begin{array}{cc}
 q& v_{n,m+1} \\ \frac{k^2}{v_{n,m}}&
 q\frac{v_{n,m+1}}{v_{n,m}}\end{array}\right)\   .
 \ee
 Using eqs. (\ref{eq:2.3}) to eliminate the backward shifts on
 $\psi_{n,m}$ we obtain an isomonodromic deformation problem in the
 usual sense. In addition, we have the linear differential equations
 with respect of the lattice parameters, namely
 \be\label{eq:MKdVpeq}
 \frac{\partial\psi_{n,m}}{\partial p} = \frac{n}{p}\left(
 \begin{array}{cc} 1&0\\ 0&0\end{array}\right) \psi_{n,m}
 + \frac{2nv_{n,m}}{v_{n+1,m}+v_{n-1,m}} \left( \begin{array}{cc}
 0&-\frac{1}{p}v_{n+1,m} \\ 0&1\end{array}\right) \psi_{n-1,m}\   ,
 \ee
 and a similar equation w.r.t. the parameter $q$.
 The infinite matrix scheme of section 2 provides us with
 a derivation of the monodromy problems as well. Details of the
 derivation can be found in Appendix A, cf. also \cite{Carg}. The
 monodromy problem (\ref{eq:MKdVmono}) was originally given for
 the special values of the parameters $\mu=0$, $\ld=0$. The present
 derivation is also taking into account boundary terms which
 are responsible for the occurrence of the free parameters.
 Our results correspond to the ones of ref. \cite{Bobenko2}, where it
 was noted that the monodromy problem has the same form as the
 one for PVI.
 
 The monodromy problem for the system in terms of $z_{n,m}$ is obtained
 from a gauge transformation on (\ref{eq:MKdVmono}), whereas the
 monodromy problem for the system in terms of $u_{n,m}$, i.e. eq.
 (\ref{eq:kdvconstr}) together with (\ref{eq:dkdv}) is given by the Lax
 pair
 \bse\label{eq:2.4}  \bea
 \phi_{n+1,m} &=& {\cal  L}_{n,m}(k^2) \phi_{n,m}\   , \label{eq:2.4a} \\
 \phi_{n,m+1} &=& {\cal  M}_{n,m}(k^2) \phi_{n,m}\   , \label{eq:2.4b}
 \eea \ese
 with Lax matrices
 \be \label{eq:KdVlax}
 {\cal L}_{n,m}(k^2)=\left(\begin{array}{cc}
 p-u_{n+1,m}&1 \\ k^2-p^2+\ast & p+u_{n,m}\end{array}
 \right) \     \ ,
 \     \
 {\cal M}_{n,m}(k^2)=\left(\begin{array}{cc}
 q-u_{n,m+1}&1 \\ k^2-q^2+\ast & q+u_{n,m}\end{array}\right)\   ,
 \ee
 in which the $\ast$ in each matrix stands for the product of the
 diagonal entries, in conjunction with the differential equation
 \bea
 k\frac{d}{dk}\phi_{n,m} &=&
 \left( \begin{array}{cc}
 \gm_{n+m}&0\\ (2\ld (-1)^{n+m}-1)(np+mq)&
 \gm_{n+m+1} \end{array} \right) \phi_{n,m} + \nn \\
 &&+\frac{2np^2}{2p+u_{n-1,m}-u_{n+1,m}} \left(
 \begin{array}{cc} -1&0\\ p-u_{n+1,m}&0\end{array} \right)
 \phi_{n-1,m} \nn \\
 && + \frac{2mq^2}{2q+u_{n,m-1}-u_{n,m+1}} \left(
 \begin{array}{cc} -1&0 \\ q-u_{n,m+1}&0 \end{array} \right)
 \phi_{n,m-1}\   .  
  \label{eq:KdVmono}
 \eea
 The compatibility of (\ref{eq:2.3}) for $\phi_{n,m}$ with the
 matrices (\ref{eq:KdVlax}) and (\ref{eq:KdVmono}) leads to the
 lattice KdV (\ref{eq:dkdv}) as well as to the similarity
 constraint (\ref{eq:kdvconstr}). In addition we have the
 following continuous deformation in terms of the dependence of
 $\phi_{n,m}$ on the lattice parameters $p$ and $q$, namely
 \be \label{eq:KdVpeq}
 \frac{\partial \phi_{n,m}}{\partial p} =
 n\left( \begin{array}{cc} 0&0\\1&0 \end{array}\right) \phi_{n,m}  \\
 + \frac{2np}{2p+u_{n-1,m}-u_{n+1,m}} \left(
 \begin{array}{cc} 1&0\\ -(p-u_{n+1,m}) &0 \end{array}
 \right) \phi_{n-1,m}\   .  \nn
 \ee
 and a similar equation in terms of $q$.
 
 Finally, a monodromy problem for the SKdV equation, in terms of the
 variable $z_{n,m}$, is easily obtained from (\ref{eq:MKdVmono}) by a
 simple gauge transformation with the matrix
 \[ \left( \begin{array}{cc} 1 & 0\\0&1/v_{n,m}\end{array}\right)   \]
 using the relations (\ref{eq:dHC}). We omit the details
 here (cf. \cite{Dorf} for the explicit form).

 \subsection{Reduced Lax Pairs}
 
 By eliminating the back-shifted vectors $\psi_{n-1,m}$
 and $\psi_{n,m-1}$ in (\ref{eq:MKdVmono}) using the
 Lax representation (\ref{eq:2.3}) we arrive at the
 following linear differential equation with regular
 singularities at $k^2=0,p^2,q^2$ and $\infty$:
 \be\label{eq:monoexpl}
 \frac{\partial\psi_{n,m}}{\partial k^2} = \left(
 \frac{S_{n,m}}{k^2} + \frac{P_{n,m}}{k^2-p^2} +
 \frac{Q_{n,m}}{k^2-q^2}\right) \psi_{n,m}
 \ee
 in which the matrices $S_{n,m}$, $P_{n,m}$ and $Q_{n,m}$
 are given by
 \begin{eqnarray*}
 &&S_{n,m}=\left(\begin{array}{cc}
 -\frac{\mu+1}{2} & \frac{n}{p}\frac{v_{n+1,m}v_{n-1,m}}
 {v_{n+1,m}+v_{n-1,m}}+\frac{m}{q}\frac{v_{n,m+1}v_{n,m-1}}
 {v_{n,m+1}+v_{n,m-1}} \\ 0 & \frac{1}{2}\gm_{n+m}
 -\frac{nv_{n-1,m}}{v_{n+1,m}+v_{n-1,m}}-\frac{mv_{n,m-1}}
 {v_{n,m+1}+v_{n,m-1}} \end{array} \right) \  , \\
 && P_{n,m}=\frac{n}{p(v_{n+1,m}+v_{n-1,m})}\left(
 \begin{array}{cc} pv_{n+1,m} & -v_{n+1,m}v_{n-1,m}\\
 -p^2 & pv_{n-1,m}\end{array} \right) \    ,\\
 &&Q_{n,m}=\frac{m}{q(v_{n,m+1}+v_{n,m-1})}\left(
 \begin{array}{cc} qv_{n,m+1} & -v_{n,m+1}v_{n,m-1}\\
 -q^2 & qv_{n,m-1}\end{array} \right) \  .
 \end{eqnarray*}
 As was noted already in \cite{Bobenko2}
 this linear differential equation has the same form as the one
 in the monodromy problem for continuous PVI, cf.
 \cite{Jimbo,JMU,JimboMiwa}. In fact, the continuous isomonodromic
 deformation is provided by the linear differential equations
 in terms of the lattice parameters, namely
 \be \label{eq:Peq}
 \frac{\partial\psi_{n,m}}{\partial p^2} =
 \left( P^0_{n,m}-\frac{P_{n,m}}{k^2-p^2}\right) \psi_{n,m}
 \ee
 where
 \[  P^0_{n,m}=\frac{n}{p(v_{n+1,m}+v_{n-1,m})}\left(
 \begin{array}{cc} \frac{1}{2p}(v_{n-1,m}-v_{n+1,m}) & 0 \\
 1&0 \end{array} \right) \   ,\]
 and a similar equation for $q$. From (\ref{eq:Peq})
 together with (\ref{eq:monoexpl}) we can derive the
 Lax pair for PVI. The {\it discrete} isomonodromic
 deformation is given by the discrete Lax pair
 (\ref{eq:2.3}) with (\ref{eq:MKdVlax}). Together with
 (\ref{eq:monoexpl}) we have a Lax pair for the
 constrained system consisting of (\ref{eq:dmkdv})
 and (\ref{eq:mkdvconstr}). To derive from it an explicit
 isomonodromic deformation problem for the GDP
 we perform the gauge transformation
 \be\label{eq:GDPgauge}
 \ol{\psi}_n=\ol{\psi}_n(m)
 \equiv \left(\begin{array}{cc} 1&0\\0&v_{n+1,m}
 \end{array}\right)\psi_{n,m} \   , \ee
 which allow us to write the monodromy entirely in terms
 of the reduced variables $x_n$, or equivalently $y_n$, of
 (\ref{eq:x-y}) which leads to
 \bse\label{eq:PVImono} \be \label{eq:PVImonoa}
 \frac{\pl\ol{\psi}_n}{\pl\kappa} =
 \left(\frac{A_n}{\kappa} +\frac{B_n}{\kappa-t}+\frac{C_n}{\kappa-s}
 \right) \ol{\psi}_n  ,
 \ee
 with $\kp=k^2$, $t=p^2$ and $s=q^2$ and where the matrices
 $A_n$, $B_n$ and $C_n$ are given by
 \bea\label{eq:PVImonob}
 A_n=\frac{1}{2}\left( \begin{array}{cc}
 -(\mu+1) & \frac{n}{p}(1-a_n)+\frac{m}{qy_n}(1-b_n) \\
 0& \gm_{n+m}-n(1-a_n)-m(1-b_n)    \end{array} \right)\   \nn \\
 B_n=\frac{n}{2}\left( \begin{array}{cc}
 1+a_n & -\frac{1}{p}(1-a_n) \\
 -p(1+a_n) & 1-a_n \end{array} \right)\    \ ,\    \
 C_n=\frac{m}{2}\left( \begin{array}{cc}
 1+b_n & -\frac{1}{qy_n}(1-b_n) \\
 -qy_n(1+b_n) & 1-b_n \end{array} \right)\    \nn \\
 \eea\ese
 where $a_n$ can be expressed in terms of $x_n$ or $y_n$
 via eqs. (\ref{eq:axy}) and $b_n$ via the similarity constraint
 ~$mb_n=\mu-\ld(-1)^{n+m}-na_n$~.
 On the one hand we can supplement (\ref{eq:PVImono}) with the
 discrete equation
 \be \label{eq:dPVIlax}
 \ol{\psi}_{n+1} = \left( \begin{array}{cc}
 p & 1 \\
 \kappa\frac{1+a_{n+1}}{1-a_{n+1}}& p\frac{1+a_{n+1}}{1-a_{n+1}}
 \end{array} \right) \psi_{n,m}\   , \ee
 leading to the eq. (\ref{eq:MPII}) or with the continuous equation
 \be\label{eq:PVIlaxt}
 \frac{\pl\ol{\psi}_n}{\pl t} =
 \frac{1}{2t}\left( \begin{array}{cc}
 -na_n & 0  \\ n\sqrt{t}\,(1+a_n)&-(n+1)a_{n+1}\end{array}\right)
 \ol{\psi}_n - \frac{B_n}{\kappa-t}\ol{\psi}_n   ,
 \ee
 By direct computation the compatibility of the above linear systems
 can be verified and the conlusions can be summarised by the following
 statement:
 
 \paragraph{Proposition 4.2:} {\it The system 
 consisting of the differential equation (\ref{eq:PVImono}) together 
 with the continuous deformation (\ref{eq:PVIlaxt})
 is compatible iff the ODE (\ref{eq:PVI?}) is satisfied leading to the 
 PVI equation. The same differential equation (\ref{eq:PVImono}) together 
 is compatible with the discrete deformation (\ref{eq:dPVIlax}) iff 
 the ordinary difference equation (\ref{eq:mpii}) is satisfied. Both the 
 discrete and the continuous deformation are isomonodromic. } 
 
 \paragraph{Remark 4.3:}
 To make the connection with the standard monodromy problem for
 PVI, in the form of refs. \cite{Jimbo,JMU}, we need to take
 a different gauge from the one in (\ref{eq:GDPgauge}), namely
 \be\label{eq:PVIgauge}
 \chi_{n,m}\equiv V_{n,m}\psi_{n,m}\    \ ,\    \
 V_{n,m}=\left(\begin{array}{cc} 1/v_{n,m}&0\\
 (u_{n,m}-np-mq)/v_{n,m}&1
 \end{array}\right) \   , \ee
 where the entry in left-lower corner of the matrix $V_{n,m}$
 can be expressed in terms of $v_{n,m}$ using the similarity
 constraint for $u_{n,m}$ in the form (\ref{eq:kdvc}).
 Using (\ref{eq:dpv}) and (\ref{eq:conteqsa}) the continuous
 isomonodromic deformation (\ref{eq:Peq}) then adopts the standard
 form
 \be\label{eq:PVIstandard}
 \frac{\pl\chi_{n,m}}{\pl t}=-\frac{W_{n,m}}{\kappa-t}\chi_{n,m}
 \      \ , \     \  W_{n,m}=V_{n,m}P_{n,m}V_{n,m}^{-1}\   .
 \ee

 \section{Schlesinger Transformations and Self-Duality}
 \setcounter{equation}{0}
 
 In this section we will derive the Schlesinger and auto-B\"acklund
 transformations for the GDP equation. This will be linked to the
 general bilinear scheme for the system.
 In fact, the derivation of the bilinear scheme in this section will
 be done from a slightly different point of view from the one in
 section 3, leading to a scheme wherein all ingredients of the theory
 will be on equal footing. We must warn that in order to arrive at
 a self-contained description, in this section we deviate slightly
 from the notations of earlier sections. Thus, the variable $X_n$ will
 not be exactly the same as the one used in (\ref{eq:odd-even}), but
 differs by a simple fractional linear transformation.

 \subsection{A first auto-B\"acklund relation}
 
 In this subsection, we shall display a first auto-B\"acklund
 transformation for the GDP which can be obtained in a straightforward
 way from the basic equations. We start
 by rewriting them here as:
 \bse\label{eq:4.1}\bea
 &&{v_{n,m+1}\over v_{n+1,m+1}}-{v_{n+1,m} \over v_{n,m}} =r( {v_{n+1,m}
 \over v_{n+1,m+1}}-{v_{n,m+1} \over
 v_{n,m}}) \label{eq:4.1a}  \\
 &&n {v_{n+1,m}-v_{n-1,m}\over {v_{n+1,m}+v_{n-1,m}}}+m
 {{v_{n,m+1}-v_{n,m-1}\over {v_{n,m+1}+v_{n,m-1}}}}=\mu-\nu\label{eq:4.1b}
 \eea\ese
 where we have assumed that the point $(m,n)$ is such that $m+n$ {is even}.
 Next we introduce
 $x_{n,m}=v_{n,m}/v_{n+1,m+1}$ and from (\ref{eq:4.1a}) we obtain
 $v_{n,m+1}/v_{n+1,m}=(rx+1)/(x+r)$. Using this ratio,
 we can express (\ref{eq:4.1b}) (upshifted once in both $m$ and $n$) in
 terms of the
 $x$'s in the following way:
 \be \label{eq:4.2}
 (n+1) {{x_{n,m}+r\over rx_{n,m}+1}-x_{n+1,m}\over {{x_{n,m}+r\over
 rx_{n,m}+1}+x_{n+1,m}}}+(m+1)
 {{rx_{n,m}+1\over x_{n,m}+r}-x_{n,m+1}\over {{rx_{n,m}+1\over
 x_{n,m}+r}+x_{n,m+1}}}=\mu-\nu\   .
 \ee
 Three more equations can be written centered around the same point $x_{n,m}$.
 \bse\label{eq:4} \bea
 && n \ {{x_{n-1,m}+r\over rx_{n-1,m}+1}-x_{n,m}\over {{x_{n-1,m}+r\over
 rx_{n-1,m}+1}+x_{n,m}}}+(m+1)
 {{rx_{n,m+1}+1\over x_{n,m+1}+r}-x_{n,m}\over {{rx_{n,m+1}+1\over
 x_{n,m+1}+r}+x_{n,m}}}=\mu+\nu\  , \label{eq:4.3}   \\
 &&(n+1){{x_{n+1,m}+r\over rx_{n+1,m}+1}-x_{n,m}\over {{x_{n+1,m}+r\over
 rx_{n+1,m}+1}+x_{n,m}}}+m\
 {{rx_{n,m-1}+1\over x_{n,m-1}+r}-x_{n,m}\over {{rx_{n,m-1}+1\over
 x_{n,m-1}+r}+x_{n,m}}}=\mu+\nu\label{eq:4.4}\   ,      \\
 &&n\ {{x_{n,m}+r\over rx_{n,m}+1}-x_{n-1,m}\over {{x_{n,m}+r\over
 rx_{n,m}+1}+x_{n-1,m}}}+ m\
 {{rx_{n,m}+1\over x_{n,m}+r}-x_{n,m-1}\over {{rx_{n,m}+1\over
 x_{n,m}+r}+x_{n,m-1}}}=\mu-\nu\label{eq:4.5} \   .
 \eea\ese
 Note that there is some asymmetry in the $m$, $n$ factors of these
 equations. This is due to the fact that
 $x_{n,m}$ is in fact defined at point $(n+1/2,m+1/2)$ whenever
 $(n,m)$ corresponds to the lattice site at which we evaluate $v_{n,m}$.
 What is the meaning of $x_{n,m}$? Clearly, this is the same $x$ as the one
 defined in Section 3, related to the
 evolution in the $n$ direction. The index $m$ here indicates simply the
 parameter value of the discrete
 equation and thus equations (\ref{eq:4.2})-(\ref{eq:4}) are just
 the auto-B\"acklunds ofthe GDP since they link
 $x_{n,m}$ and $x_{n+1,m}$ (or $x_{n-1,m}$) to $x_{n,m+1}$.
 These auto-B\"acklund transformations do not cover all the possibilities. In
 particular, the changes of the
 parameters $\nu$ and $\mu$ can also intoduce Schlesinger transformations.
 The latter will be examined in the
 next subsection in the framework of the self-dual approach.
 
 A last interesting feature of the two-dimensional equation
 (\ref{eq:4.2})-(\ref{eq:4}) is that
 they can be used in order to define
 one dimensional evolutions in direction different from those of the
 orthogonal lattice $(m,n)$. Let us consider
 for example the system of equations (\ref{eq:4.3}) together with the
 $n$-downshifted of equation (\ref{eq:4.4}).
 \be\label{eq:4.59}
 n{{x_{n,m}+r\over rx_{n,m}+1}-x_{n-1,m}\over {{x_{n,m}+r\over
 rx_{n,m}+1}+x_{n-1,m}}}+m{{rx_{n-1,m-1}+1\over
 x_{n-1,m-1}+r}-x_{n-1,m}\over {{rx_{n-1,m-1}+1\over
 x_{n-1,m-1}+r}+x_{n-1,m}}}=\mu-\nu\    ,
 \ee
 (notice the change of sign in the r.h.s.).
 
 Introducing  the one-dimensional notation $X_n=x_{n,n+k}$,
 $Y_n=x_{n-1,n+k}$ (with $k=m_0-n_0$, where the index 0
 characterizes the starting point), we obtain thus what we call an
 asymmetric form (the
 name has its origin at the asymmetric family of QRT mappings
 introduced in \cite{QRT}, cf. also \cite{GORS}).
 \bse\label{eq:4.7}\bea
 &&(\zeta_n-\kappa) \ {{Y_{n}+r\over rY_{n}+1}-X_{n}\over
 {{Y_{n}+r\over rY_{n}+1}+X_{n}}}
 +(\zeta_{n+1}+\kappa){{rY_{n+1}+1\over Y_{n+1}+r}-X_{n}\over
 {{rY_{n+1}+1\over Y_{n+1}+r}+X_{n}}}=\mu+\nu\   , \label{eq:4.7a}   \\
 &&(\zeta_{n}-\kappa){{X_{n}+r\over rX_{n}+1}-Y_{n}\over
 {{X_{n}+r\over rX_{n}+1}+Y_{n}}}
 +(\zeta_{n}+\kappa){{rX_{n-1}+1\over X_{n-1}+r}-Y_{n}\over
 {{rX_{n-1}+1\over X_{n-1}+r}+Y_{n}}}=\mu-\nu\   ,  \label{eq:4.7b}
 \eea\ese
 with again $\zeta_n=\delta n+\zeta_0$ and $2\kappa=\delta
 k$. This mapping is a discrete form of P$_{\rm V}$ equation as
 can be assessed through the continuous limits. Indeed, putting
 $X=1+\epsilon (w+1)/(2-2w)$, $Y=1+\epsilon u$, $\delta=\epsilon$,
 $r=-(1+\epsilon/4)$, $\mu=\epsilon g$, $\zeta_0=0$,
 $\nu=\epsilon h$ we obtain:
 \be\label{eq:PV}
 w''=w'^2({1\over 2w}+{1\over w-1})-{w'\over z}+{(w-1)^2\over z^2}(a w+{b
 \over w})+{c w\over z} +{d w(w+1)\over w-1}\   ,
 \ee
 i.e. P$_{\rm V}$ with $z=n\epsilon$, $a=(k+g-h)^2/8$, $b=-(k-g+h)^2/8$,
 $c=(g+h)/2$ and $d=-1/8$.
 
 At this point it is interesting to show that the equations by
 (\ref{eq:4.7}) are equivalent to another discrete form of
 P$_{\rm V}$, obtained in \cite{GORS}.  We start by introducing the
 new variables through the following change:
 \bse\label{eq:Change}\bea
 Y_n&=&-\frac{r\Ups_n+1}{\Ups_n+r}\  , \label{eq:5.11}\\
 X_n&=&-\Ups_{n+1}
 \frac{2\Xi_n+\delta+2\kappa-\mu-\nu}{2\Xi_n-4\zeta-3\delta-2\kappa-\mu
 -\nu}\   .  \label{eq:5.12}
 \eea\ese
 {}From equations (\ref{eq:4.7}) we obtain, in terms
 of the variables $\Xi_n,\Ups_n$:
 \bse\label{eq:NewPV}\bea
 \Ups_n\Ups_{n+1}&=&\frac{(2\Xi_n-4\zeta_n-3\delta-2\kappa-\mu-\nu)
 (2\Xi_n-4\zeta_n-\delta+ 2\kappa+\mu+\nu)}
 {(2\Xi_n+\delta+2\kappa-\mu-\nu)(2\Xi_n-\delta-2\kappa+\mu+\nu)}
 \label{eq:5.121}\\
 \Xi_{n-1}+\Xi_n&=&
 \frac{2\zeta_n+\mu-\nu}{\Ups_n/r+1}+\frac{2\zeta_n+\nu-\mu}{\Ups_n+r}\   .
 \label{eq:5.13}
 \eea\ese
 Introducing the quantities $a=\mu-\nu$,
 $b=\kappa+(\delta-\mu-\nu)/2$ and
 $c=\kappa+(\delta+\mu+\nu)/2$ we can write the equations as:
 \bse\label{eq:NewdPV}\bea
 \Ups_n\Ups_{n+1}&=&\frac{(\Xi_n-\zeta_n-\zeta_{n+1}-c)
 ((\Xi_n-\zeta_n-\zeta_{n+1}+c))}{(\Xi_n+b)(\Xi_n-b)}
 \label{eq:5.14}\\
 \Xi_{n-1}+\Xi_n&=& \frac{2\zeta_n+a}{\Ups_n/r+1}+
 \frac{2\zeta_n-a}{\Ups_n+r}\label{eq:5.15}
 \eea\ese
 This system coincides precisely with the one obtained in \cite{GORS}
 from the degeneration of $q$-P$_{\rm
 VI}$. It goes without saying that the continuous limit of
 (\ref{eq:NewdPV}) is again P$_{\rm V}$.

 \subsection{Bilinear expressions and Hirota-Miwa equations}
 
 The system of equations (\ref{eq:4.2})-(\ref{eq:4}) offers a fully
 symmetric description of the evolution along the direction $n$ and
 $m$. However, the GDP has clearly more Schlesinger
 transformations (those corresponding to the
 parameters $\mu$, $\nu$) which do not seem to fit into this symmetric
 scheme. We shall show in this subsection that it is possible to
 introduce a fully self-dual description of the GDP in which all four
 directions $n$, $m$, $\mu$, $\nu$ play the same role. The key to this
 approach is the bilinear formalism.
 Let us start by the bilinearization of equations
 (\ref{eq:4.2}) and (\ref{eq:4}). We will find it useful to
 distinguish between four different $\tau$-functions will be necessary:
 $F$, $G$ that exist only for `even' lattice points (i.e. $n+m$ even) and
 $H$, $K$ present at the `odd' points, namely
 \be
 v_{n,m}={F_{n,m}\over G_{n,m}} \quad {\rm and} \quad
 v_{n+1,m}={H_{n+1,m}\over K_{n+1,m}}\   . \label{eq:4.8}
 \ee
 Thus, comparing with (\ref{eq:v0}) the $F$ and $H$ can be
 identified with the $\tau_+$ at those sites and the $G$ and
 $K$ with the $\tau_-$ (or, equivalently, the $\sg_{n,m}$.
 Next, we substitute into (\ref{eq:dmkdv}) and obtain:
 \bea && H_{n,m+1}K_{n+1,m}(pF_{n,m} G_{n+1,m+1}+qF_{n+1,m+1}
 G_{n,m})  \nn \\
 && ~~ = H_{n+1,m}K_{n,m+1}(qF_{n,m}G_{n+1,m+1}+pF_{n+1,m+1} G_{n,m})
 \label{eq:4.9} \eea
 For this expression we see that we can separate provided we add the same
 quantity
 $-(p+q)(H_{n+1,m}K_{n,m+1}H_{n,m+1}K_{n+1,m})$ to both sides of the
 equation. (The choice of the multiplicative
 factor is based on the convention of Hirota who takes the sum of the
 coefficients of the Hirota-Miwa equation to
 be equal to zero). Separating (\ref{eq:4.9}) we find.
 \bse\label{eq:4.10}\bea
 &&pF_{n,m} G_{n+1,m+1}+qF_{n+1,m+1} G_{n,m}= (p+q)H_{n+1,m}K_{n,m+1}
 \label{eq:4.10a}  \\
 && pF_{n,m} G_{n-1,m-1}+qF_{n-1,m-1} G_{n,m}= (p+q)H_{n-1,m}K_{n,m-1}
 \label{eq:4.10b}  \eea\ese
 where (\ref{eq:4.10b}) is in fact the down-shift in both $m$ and
 $n$ of what we obtain in separating (\ref{eq:4.9}). In eq.
 (\ref{eq:4.10a}) we recognise immediately (\ref{eq:addrel2a}).
 
 Thus, the autonomous mKdV equation is bilinearized to a system of
 Hirota-Miwa equations. A close look at (\ref{eq:4.10a}) and (\ref{eq:4.10b})
 reveals that the two equations, although non strictly identical, describe
 the same
 evolution. Simply the direction of evolution is reversed between
 (\ref{eq:4.10a})
 and (\ref{eq:4.10b}).
 
 Next, we turn to the  bilinearization of (\ref{eq:mkdvconstr}). Since the
 equation relates
 $v$'s at `odd' points of the lattice
 (shifted by one with respect to $v_{n,m}$) only the $H$, $K$ functions
 enter. We find:
 \bea\label{eq:4.11}
 && H_{n+1,m}K_{n-1,m}((n+m-\mu+\nu)H_{n,m+1}K_{n,m-1}+(n-m-\mu+\nu)
 H_{n,m-1}K_{n,m+1})  \nn \\
 && ~~ =H_{n-1,m}K_{n+1,m}((n-m+\mu-\nu)H_{n,m+1}K_{n,m-1}+
 (n+m+\mu-\nu)H_{n,m-1}K_{n,m +1}) \nn \\
 \eea
 Again we separate by adding the quantity
 $gH_{n-1,m}K_{n+1,m}H_{n+1,m}K_{n-1,m}$ to both sides of the equation.
 This time it is not convenient to implement the Hirota convention and
 thus we just choose $g=-2n$.
 \bse \label{eq:4.12} \bea
 &&(n+m-\mu+\nu)H_{n,m+1}K_{n,m-1}+(n-m-\mu+\nu)H_{n,m-1}K_{n,m+1}-2nH_{n-1,m}K_
 {
 n+1,m}=0 \nn \\
 \label{eq:4.12a}   \\
 &&(n-m+\mu-\nu)H_{n,m+1}K_{n,m-1}+(n+m+\mu-\nu)H_{n,m-1}K_{n,m+1}-2nH_{n+1,m}K_
 {
 n-1,m}=0 \nn \\
 \label{eq:4.12b}
 \eea\ese
 (Notice that both equation are invariant with respect to the parity $n\to
 -n$).
 Again the two equations, although not identical, describe the same
 evolution with the direction reversed.
 Expressions (\ref{eq:4.12a}) and (\ref{eq:4.12b}) are reductions of a more
 general Miwa
 equation which involves all four products
 $H_{n+1,m}K_{n-1,m}$, $H_{n-1,m}K_{n+1,m}$, $H_{n,m-1}K_{n,m+1}$ and
 $H_{n,m+1}K_{n,m-1}$ and has a free
 coefficient (which allows the reduction to be constructed).
 
 Since we already presented the bilinear form of GDP in section 3
 it makes sense to compare it to the one obtained here. We have
 used various notations for the $\tau$-functions so let us first
 establish the equivalence between them.  Comparing
 (\ref{eq:v0}) as it was reinterpreted in Section 3.1 with
 (\ref{eq:4.8}) we have:
 $\tau_{n,m}=G_{n,m},\sigma_{n,m}=F_{n,m}$ for $n+m$ even, and
 $\tau_{n,m}=K_{n,m},\sigma_{n,m}=H_{n,m}$ for $n+m$ odd.
 
 It is straightforward to show that (\ref{eq:4.12}) comes directly
 from (\ref{eq:tausim}) provided we express $\sigma_{n,m}\tau_{n,m}$
 using (\ref{eq:addrel}). Similarly (\ref{eq:4.10}) can be obtained
 from (\ref{eq:Taueqs})-(\ref{eq:addrel}) after some algebraic
 manipulations. Conversely, we can obtain (\ref{eq:Taueqs})-
 (\ref{eq:tausim}) starting from (\ref{eq:4.10}) and
 (\ref{eq:4.12}). We start by noting that both sets of equations
 are invariant under the gauge $\tau\to e^{c(n^2+m^2)}\tau$ where
 $\tau$ stands for any of the $F,G,H,K$.
 On the other hand the l.h.s. and the r.h.s. of (\ref{eq:addrel})
 behave differently under this gauge. Thus, we can use it in
 order to ensure that (\ref{eq:addrela}), for instance, is
 satisfied at a {\it given point} $(n,m)$. Once (\ref{eq:addrela})
 is satisfied at {\it one} point one can use (\ref{eq:4.10}) and
 (\ref{eq:4.12}) to prove that (\ref{eq:Taueqs}), (\ref{eq:addrel})
 and (\ref{eq:tausim}) hold true at each point of the
 two-dimensional lattice.

 \subsection{Self-dual formulation of the GDP}
 
 In the previous subsection we have seen how the GDP can
 be expressed as a system of bilinear
 autonomous and non-autonomous Hirota-Miwa equations. Although offering a
 complete description of the system at
 hand these equations treat in a completely asymmetrical way the dynamic
 variables ($n,m$) on the one hand and
 ($\mu,\nu$) on the other. However the latter play precisely the same role
 as the former since their evolution is
 also related to Schlesinger transformation. What is needed is a
 symmetrical, {\it self-dual}  treatment of all four
 parameters, in the spirit of the ``Grand Scheme'' of \cite{ROSG}.
 
 As a first step towards this self-dual description, we introduce a change
 in the notations. The
 variables $n$, $m$, $\mu$ and $\nu$ will in what follows be represented by
 $n_1$, $n_2$, $n_3$ and $n_4$. Next,
 instead of the four $\tau$-functions $F$, $G$, $H$ and $K$, we introduce a
 single function $\tau$ which is a
 function of all four variables, $\tau=\tau(n_1,n_2,n_3,n_4)$. Finally
 we use the short-hand notation introduced in \cite{ROSG}, where upper
 and lower indices are used in order to denote shifts
 $\tau^k=\tau(\dots,n_k+1,\dots)$, $\tau_k=\tau(\dots,n_k-1,\dots)$,
 (with the obvious disadvantage that a shift of $j$ steps requires $j$
 repetitions of the index of the respective variables).
 
 {}From the results of the previous subsection 5.2, cf. also subsection
 2.3, we know that $v$  is a ratio of two $\tau$'s and it turns out
 that the good choice is to consider only $\tau$-functions with an
 even number of shifts.  We have thus the following identification of
 the $\tau$ functions of the previous subsection (it goes without
 saying that there is some arbitrariness in the choice of this
 correspondence):
 \be\label{eq:4.13} F_{n,m}\equiv\tau^{34},\quad G_{n,m}\equiv\tau,\quad
 H_{n+1,m}\equiv\tau^{13},\quad K_{n+1,m}\equiv\tau^{14}\   . \ee
 Note that for the $\tau$-functions $F$ and $G$ the sum of the shifts
 in the directions 1  or 2 is even, while this sum is odd for $H$ and
 $K$.  We can now transcribe equations (\ref{eq:4.10}) to:
 \bse\label{eq:4.14}\bea
 &&p\tau^{34} \tau^{12}+q\tau^{1234}\tau=
 (p+q)\tau^{13}\tau^{24}\   ,
 \label{eq:4.14a}   \\
 &&p\tau^{34} \tau_{12}+q\tau_{12}^{34}\tau= (p+q)\tau_{1}^{3}\tau_{2}^{4}
 \label{eq:4.14b}
 \eea\ese
 Similarly, the non-autonomous equations (\ref{eq:4.12}) are written as:
 \bse\label{eq:4.15}\bea
 &&(n_1+n_2-n_3+n_4)\tau^{23}\tau_{2}^{4}+(n_1-n_2-n_3+n_4)\tau_{2}^{3}
 \tau^{24}-2n_1\tau_{1}^{3}\tau^{14} =0\label{eq:4.15a}   \\
 &&(n_1-n_2+n_3-n_4)\tau^{23}\tau_{2}^{4}+(n_1+n_2+n_3-n_4)\tau_{2}^{3}
 \tau^{24}-2n_1\tau^{13}\tau_{1}^{4} =0\label{eq:4.15b}
 \eea\ese
 This system is not sufficient in order to describe the evolution in all
 four directions. We can in principle
 extend these particular forms formally to the remaining dimensions by
 permuting  the indices, but this leads to
 incompatibilities. So we must first modify (\ref{eq:4.14}) and
 (\ref{eq:4.15}),
 introducing the appropriate gauge factor $\tau\to\phi\tau$. To make a
 (very) long story short we give here
 the factor $\phi$
 \bea\label{eq:4.16}
 \phi=&&\left({a_1-a_4 \over a_1+a_4}\right)^{n_1n_4/2}\!\!\left({a_2-a_4 \over
 a_2+a_4}\right)^{n_2n_4/2}
 \!\!\left({a_3-a_1 \over a_3+a_1}\right)^{n_1n_3/2}\left({a_3-a_2 \over
 a_3+a_2}\right)^{n_2n_3/2}
 \!\!\left({a_3-a_4 \over a_3+a_4}\right)^{n_3n_4/2} \nn \\
 && \left({a_3^2-a_1^2 \over a_1a_3}\right)^{n_1^2+n_3^2\over 4}
 \left({a_3^2-a_2^2 \over a_2a_3}\right)^{n_2^2+n_3^2\over 4}
 \left({a_1^2-a_4^2 \over a_1a_4}\right)^{n_1^2+n_4^2\over 4}
 \left({a_2^2-a_4^2 \over a_2a_4}\right)^{n_2^2+n_4^2\over
 4}\Omega^{n_1n_2/2}
 \eea
 with:
 \be \Omega={a_1-a_2\over
 a_1+a_2}{\sqrt{a_1^2-a_4^2}\sqrt{a_3^2-a_2^2}+\sqrt{a_2^2-a_4^2}\sqrt{a_3^2-a_1
 ^
 2}
 \over
 \sqrt{a_1^2-a_4^2}\sqrt{a_3^2-a_2^2}-\sqrt{a_2^2-a_4^2}
 \sqrt{a_3^2-a_1^2}}\  ,  \label{eq:agauge}\ee
 where the $a_i$, $i=1,2,3,4$, are any four numbers subject only to
 the constraint
 \be\label{eq:aconstr}
 \frac{p}{q}=\frac{\sqrt{a_2^2-a_4^2}}{\sqrt{a_1^2-a_4^2}}
 \frac{\sqrt{a_3^2-a_1^2}}{\sqrt{a_3^2-a_2^2}}.
 \ee
 The careful application of this transformation leads to the system
 consisting of:
 \bse\label{eq:Grand}\bea
 && (a_1-a_2)(a_3-a_4)\tau^{12}\tau^{34}+(a_1-a_3)(a_4-a_2)\tau^{13}\tau^{24}+
 (a_1-a_4)(a_2-a_3)\tau^{14}\tau^{23}=0\    , \nn \\
 \label{eq:4.17}  \\
 && 2n_1a_2(a_1-a_3)(a_1-a_4)\tau_{13}\tau^{14} +
 (n_1+n_2+n_3+n_4)a_1(a_2+a_4)(a_2+a_3)\tau_2^{4}\tau_3^{2}
 \nn  \\
 && ~~~~~ -(n_1-n_2+n_3+n_4)a_1(a_2-a_4)(a_2-a_3)\tau_{23}\tau^{24}
 =0\   ,
 \label{eq:4.18}
 \eea\ese
 which must be understood not only as they stand, but also with all
 possible permutations of  $\{ 1,2,3,4\}$ and
 also any number of the parities
 $\tau^{k}\to\tau_{k}$, $a_k\to-a_k$, $n_k\to-n_k$. Equations
 (\ref{eq:4.17}) and (\ref{eq:4.18}) (augmented as we just explained) are
 obviously self-dual. All four directions play the same role. Thus, the
 GDP can be understood as defining an evolution in a four-dimensional
 discrete space ${\bf Z}^4$.

 \subsection{The remaining auto-B\"acklund transformations}
 
 In order to introduce the auto-B\"acklund transformations for the
 remaining directions, ($\mu,\nu$) we
 start with some considerations of singularity structure. Using the bilinear
 formalism introduced in the preceding subsection, we have:
 \be x={v\over v^{12}}={\tau^{12} \tau^{34}\over \tau
 \tau^{1234}}\label{eq:4.19}
 \ee
 The second special expression involving $x$ is also simply given as
 \be
 {x+r\over rx+1}={v^1\over v^{2}}={\tau^{13} \tau^{24}\over \tau^{14}
 \tau^{23}}\label{eq:4.20}
 \ee
 where $r=q/p$.
 {}From these two expressions, one concludes immediately that the four
 singular values of $x$ are $0$, $\infty$,
 $-r$ and $-1/r$. As we have seen in subsection 4.1 the auto-B\"acklund
 transformations between directions 1 and 2 involve
 directly expressions (\ref{eq:4.19}) and (\ref{eq:4.20}).
 Let us now consider the auto-B\"acklund transformations that relate
 directions 1 and 3, i.e. the Schlesinger
 transformation for $n_3$ while the evolution is still along $n_1$.
 Permuting 2 and 3 in expressions (\ref{eq:4.19})  we
 find that, {\sl up to a gauge}, the relevant quantity is:
 \be y={\tau^{13} \tau^{24}\over \tau \tau^{1234}}\label{eq:4.21}  \ee
 One can understand the origin of this gauge factor in the following way.
 When starting from the fully self-dual
 equations, one must introduce a {\sl different} gauge whether one wants to
 specify an evolution in the couple of
 directions (1,2) or (1,3). Thus the relative gauge of $y$ with respect to
 $x$ comes from the combination of these
 two gauges.
 Comparing (\ref{eq:4.21}) and  (\ref{eq:4.19}), and using
 (\ref{eq:4.14}),  we conclude that $y\propto (x+r)$ with a
 multiplicative factor that remains to be
 fixed by the appropriate gauge. Instead of computing this factor by
 carefully going through the gauge
 transformations we use a short-cut.
 We start by remarking that the permutation of indices corresponds to the
 permutation of the singularities. Let
 us give an example. As we have seen, the permutation of indices 2 and 3
 transforms $x$ to $x+r$ (up to a global
 factor). Thus, the singularities $\{0,\infty,-r,-1/r\}$ are transformed
 into $\{r,\infty,0,r-1/r\}$ but we know
 that, in the appropriate gauge, the product of the two finite singularities
 must be unity. So we must renormalize
 them by a factor $\lambda=1/\sqrt{r^2-1}$, which leads to  $y=
 (x+r)/\sqrt{r^2-1}$ having its  singularities at
 $\{-s,\infty,0,-1/s\}$ with $s=-r/\sqrt{r^2-1}$.

 We can now establish the auto-B\"acklund transformations involving
 direction 3. We find:
 \be
 (n+1) {{y_{n,\mu}+s\over sy_{n,\mu}+1}-y_{n+1,\mu}\over
 {{y_{n,\mu}+s\over sy_{n,\mu}+1}+y_{n+1,\mu}}}+(\mu+1)
 {{sy_{n,\mu}+1\over y_{n,\mu}+s}-y_{n,\mu+1}\over {{sy_{n,\mu}+1\over
 y_{n,\mu}+s}+y_{n,\mu+1}}}=m-\nu\   , \label{eq:4.22}
 \ee
 which is the equivalent of equation (\ref{eq:4.2}) of subsection 4.1. The
 analogs of
 equations (\ref{eq:4}) can be established in a straightforward way.
 The auto-B\"acklund relations involving direction 4 can be obtained in a
 similar way.
 
 \section{Conclusions}
 
 In this paper we have established a direct connection between
 the fully discretised lattice systems associated with the KdV
 family and Painlev\'e equations via the similarity reduction
 on the lattice. In particular, under the reduction PVI emerges as
 the continuous equation in terms of the lattice {\it parameters}.
 It is well-known that Schlesinger transformations act on the solutions
 of PVI in a discrete way , cf. e.g. \cite{Mugan}, and it is therefore
 not surprising
 that a lattice structure sits at the background of this equation.
 However, by making the connection with the lattice KdV we have
 been able to identify this lattice structure in explicit terms, and
 in its most attractive form it leads to an explicit difference
 equation, the GDP (\ref{eq:MPII}). This equation, as an
 integrable difference equation of ``Painlev\'e type'', can then be
 studied in its own right. Furthermore, the relevant bilinear
 structure of PVI, found in \cite{ORGT}, seems to fit naturally in the
 discrete structure demonstrating that, in fact, the continuous and
 discrete aspects for these systems are intimately interlaced.
 The self-dual bilinear structure of the discrete equations exhibited
 in subsection 5.3 demonstrates that independent variables and
 parameters of the Painlev\'e equations live on equal footing.
 
 We expect that the results highlighted in this paper will allow
 to gain new insights in the nature of both the discrete as well
 as the continuous transcendents.
 The rather complicated combinatorics that is revealed
 by some of the derivations, e.g. in subsection 3.1 and in the
 patterns of section 4.1, we believe are indicative of the
 existence of some highly nontrivial addition formulae
 for the transcendental solutions, reminiscent of the contiguous
 relations for functions of hypergeometric type. Ultimately, we
 hope that these results will form an inspiration for a future
 analytic theory of ordinary difference equations.

 \pagebreak
 
 \def\thesection{Appendix A:}
 
 \section{Derivation of Monodromy Problems}
 \setcounter{equation}{0}
 \def\theequation{A.\arabic{equation}}
 
 We present here some of the details in the derivation of the
 monodromy problems for the GDP, (\ref{eq:MPII}), which was also
 given in the form (\ref{eq:mpii}). 
 We will derive the relevant formulae from the infinite matrix
 structure of section 2. All Lax pairs and monodromy problems can be
 derived from that structure.
 
 \paragraph{a) Lattice (potential) KdV:} In this case we take the
 components ~$u_k^0$~ and ~$u_k^1$~ to form the
 two-component vector
 \be \phi_k\equiv (p-k)^n(q-k)^m \left( \begin{array}{c}
 u_k^0 \\ u_k^1 \end{array}\right)\   ,
 \ee
 for which it is immediate to derive the linear relation
 \be\label{eq:kdvlax}
 \wt{\phi}_k = L_k\phi_k\    \ ,\      \
 L_k\equiv \left( \begin{array}{cc}
 p-\wt{u}& 1\\ k^2-p^2 + \ast & p+u\end{array}\right)\   ,
 \ee
 in which $u={\bf U}_{0,0}$ as before and $\ast$ stands for the product
 of the diagonal entries of the matrix $L_k$. Eq. (\ref{eq:kdvlax})
 is one part of the Lax pair of the potential lattice KdV equation.
 The other part is identical apart from the replacements:
 $\wt{u}\mapsto \widehat{u}$, $p\mapsto q$, leading to
 \be\label{eq:kdvlax2}
 \widehat{\phi}_k = M_k\phi_k\    \ ,\      \
 M_k\equiv \left( \begin{array}{cc}
 q-\widehat{u}& 1\\ k^2-q^2 + \ast & q+u\end{array}\right)\   .
 \ee
 In this way we obtain the Lax representation for (\ref{eq:dkdv}) from the
 compatibility condition
 \[ \widehat{L}_k M_k=\wt{M}_kL_k\   . \]
 In order to obtain the monodromy problem that gives us the similarity
 constraint we have to use (\ref{eq:monouk}) and write down what this
 gives us for the vector $\phi_k$. Applying the constraint for
 $u^0_k$ and $u^1_k$ respectively we obtain
 \bse\bea\label{eq:uk0}
 k\frac{d}{dk}u_k^0 &=& \ld(-1)^{n+m} u_k^0 +
 np\left( \frac{1}{p-k}u_k^0 -v_{-p}u_k^{(p)}\right)
 + mq\left( \frac{1}{q-k}u_k^0 -v_{-q}u_k^{(q)}\right)\   , \nn \\
 \\
 k\frac{d}{dk}u_k^1 &=& \left( 1 - \ld(-1)^{n+m}\right) u_k^1
 + 2\ld(-1)^{n+m}(pn+mq) u_k^0  \nn \\
 && ~~ + np\left( \frac{1}{p-k}u_k^1 - u_k^0 - s_{-p}u_k^{(p)}\right)
 + mq\left( \frac{1}{q-k}u_k^1 - u_k^0 - s_{-q}u_k^{(q)}\right)\   ,
 \nn \\  \label{eq:uk1} \eea\ese
 where $u_k^{(\ar)}$ is defined by
 \be\label{eq:ukar}
 u_k^{(\ar)}\equiv \left(\frac{1}{\ar+\Ld}\cdot {\bf u}_k\right)_0\   ,
 \ee
 and where the objects $v_\ar$ and $s_\ar$ were defined in
 (\ref{eq:vs}). In order to bring this set of equations into shape
 we need a number
 of relations, namely (\ref{eq:relst}) and (\ref{eq:miura}) for
 $\ar=-p,-q$ together with (\ref{eq:st}) and (\ref{eq:vv}),
 as well as from (\ref{eq:linuk}) the relations
 \be\label{eq:pukar}
 (p-k)\wt{u}_k^{(\ar)} = (p-\ar) u_k^{(\ar)} + \wt{v}_\ar u_k^0\   ,
 \ee
 applied to $\ar=p$, (and similarly to the equations with $q$).
 
 For the dependance on the B\"acklund parameters $p$ and $q$ we have the
 following relations
 \bse\bea\label{eq:puk0}
 \frac{\pl}{\pl p}u_k^0 &=& n \left( v_{-p} u_k^{(p)} -
 \frac{1}{p-k} u^0_k \right)\   ,    \\
 \frac{\pl}{\pl p}u_k^1 &=& n\left( u_k^0 + s_{-p} u_k^{(p)} -
 \frac{1}{p-k} u_k^1 \right)\   ,  \label{eq:puk1}
 \eea\ese
 with similar relations describing the dependence  on $q$.
 Thus, we obtain a linear system of the following type
 \be
 \frac{\pl \phi_k}{\pl p}= n \left(
 \begin{array}{cc} 0 & 0\\ 1&0
 \end{array} \right) \phi_k+ n \left( \begin{array}{cc}
 v_{-p}v_p & 0 \\  s_{-p}v_p & 0
 \end{array} \right) {\hypotilde 0 \phi_k}\    ,
 \ee
 and similarly a linear system in terms of $q$.
 
 The result is the following differential equation for the vector
 $\phi_k$
 \bea
 && k\frac{d}{dk}\phi_k = \left( \begin{array}{cc}
 n+m&0\\ -np-mq&n+m+1\end{array} \right) \phi_k
 + \ld(-1)^{n+m} \left( \begin{array}{cc}
 1&0\\ 2np+2qm&-1\end{array} \right) \phi_k \nn \\
 && ~~ -\frac{2np^2}{2p+\ut{u}-\wt{u}} \left(
 \begin{array}{cc} 1 & 0 \\ -p+\wt{u}&0 \end{array} \right)
 {\hypotilde 0 \phi}_k
 -\frac{2nq^2}{2p+{\hypohat 0 u}-\wh{u}} \left(
 \begin{array}{cc} 1 & 0 \\ -q+\wh{u}&0 \end{array} \right)
 {\hypohat 0 \phi}_k\   ,   \label{eq:umono}
 \eea
 which is (\ref{eq:KdVmono}. We must next use the inverse of the
 Lax representation (\ref{eq:2.3}), (\ref{eq:2.4}) to express
 ${\hypotilde 0 \phi}_k$ and ${\hypohat 0 \phi}_k$ in terms of
 $\phi_k$, thus yielding a pure linear differential equation
 in terms of the spectral parameter $k^2$.
 
 \paragraph{b) Lattice (potential) MKdV:}
 To obtain the Lax representation for the lattice MKdV equation a similar
 calculation need to be performed. In this case we single out the components
 $u^0_k$ together with $u_k^{(\ar)}$ for some fixed value of $\ar$, to
 construct the two-component vector
 \be
 \psi_k\equiv (p-k)^n(q-k)^m\left( \begin{array}{c}
 u_k^{(\ar)}\\ u_k^0 \end{array}\right)\   .
 \ee
 First, using the relations (\ref{eq:relst}) we obtain from
 (\ref{eq:monouk})
 now the Lax matrices in the form
 \be\label{eq:mkdvlax}
 \wt{\psi}_k = L_k\phi_k\    \ ,\      \
 L_k\equiv \left( \begin{array}{cc}
 p-\ar& \wt{v}_\ar \\ \frac{k^2-\ar^2}{v_\ar}&
 (p+\ar) \frac{\wt{v}_\ar}{v_\ar}\end{array}\right)\   ,
 \ee
 and a similar expression for $M_k$ by making the usual replacements.
 
 \paragraph{Remark A.1:} We may use eqs. (\ref{eq:relst}) to show
 that the Lax representations
 (\ref{eq:kdvlax}) and (\ref{eq:mkdvlax}) are related via the
 following gauge transformation, 
 \be\label{eq:gauge}
 L_k^{(MKDV)} = \wt{U}_k L_k^{(KDV)} U_k^{-1}\    \ ,\    \
 U_k=\left( \begin{array}{cc} -s_\ar & v_\ar \\
 k^2-\ar^2 & 0\end{array}\right)\   ,
 \ee
 (and a similar formula relating the matrices $M_k$).
 
 Now, the dependence on the B\"acklund parameters $p$ and $q$ of the quantity
 $u_k^{(\ar)}$ plays a role, and this is given by the relation
 \be\label{eq:puka}
 \frac{\partial}{\partial p}u_k^{(\ar)} = n \left[
 \frac{\ar-k}{(p-\ar)(p-k)} u_k^{(\ar)}
 - \left( s_{\ar,-p} + \frac{1}{p-\ar}\right) u_k^{(p)}\right]\   ,
 \ee
 with a similar relation describing the dependence  on $q$. Thus, we get
 for the vector $\psi_k$ for $\ar=0$ the following linear system
 \be
 p\frac{\pl\psi_k}{\pl p}= n\left( \begin{array}{cc} 1 & 0\\ 0&0
 \end{array} \right) \psi_k+ n \left( \begin{array}{cc} 0 &
 -(1+ps_{0,-p})v_p\\
 0&pv_{-p}v_p  \end{array} \right) {\hypotilde 0 \psi_k}\    ,
 \ee
 and an accompanying linear system in terms of the dependence on $q$.

 In order to derive a monodromy problem from (\ref{eq:monouk}) we need to
 restrict ourselves to the case $\ar=0$, because otherwise we would obtain
 from the term with the operator ${\bf I}$ derivatives with respect to
 $\ar$, hence equations which are no longer closed-term. Using the
 following additional relation for $u_k^{(0)}\equiv u_k^{(\ar=0)}$
 \bea
 k\frac{d}{dk}u_k^{(0)} &=& -(\mu+1)u_k^{(0)} +
 n\left( \frac{k}{p-k}u_k^{(0)} + (1+ps_{0,-p}) u_k^{(p)}\right)  \nn \\
 &&  + m\left( \frac{k}{q-k}u_k^{(0)} + (1+qs_{0,-q})u_k^{(q)}\right)\   ,
 \label{eq:uk-1}
 \eea
 in addition to (\ref{eq:uk0}) and the relations (\ref{eq:sta})
 and (\ref{eq:stc}) we easily obtain the following monodromy problem
 \bea
  k\frac{d}{dk}\psi_k &&= \left( \begin{array}{cc}
 -(\mu+1)&0\\ 0&\ld (-1)^{n+m}\end{array} \right) \psi_k
  + (n+m)\left( \begin{array}{cc}
 0&0\\ 0&1\end{array} \right) \psi_k
 \label{eq:vmono} \\
 && ~~ +\frac{2nv_0}{\wt{v}_0+\ut{v_0}} \left(
 \begin{array}{cc} 0 & \wt{v}_0 \\  0& -p \end{array} \right)
 {\hypotilde 0 \psi}_k
 +\frac{2mv_0}{\wh{v}_0+ {\hypohat 0 v_0}} \left(
 \begin{array}{cc} 0& \wh{v}_0  \\ 0& -q \end{array} \right)
 {\hypohat 0 \psi}_k\   ,   \nn
 \eea
 which is (\ref{eq:MKdVmono}). Again, in order to obtain the
 purely differential equation we need to use the inverse of the
 discrete Lax pair (\ref{eq:mkdvlax}), and its counterpart,
 to replace the backward-shifted ${\hypotilde 0 \psi}_k$ and
 ${\hypohat 0 \psi}_k$.

 \pagebreak

 \end{document}